\documentclass[a4paper,11pt]{article}
\pdfoutput=1 
\usepackage{jcappub}

\usepackage{subcaption}
\usepackage{hyperref}
\usepackage[section]{placeins}
\usepackage[titletoc]{appendix}
\usepackage{xcolor}
\usepackage{dsfont}
\usepackage[T1]{fontenc}
\usepackage[utf8]{inputenc}
\usepackage{bm}
\usepackage{mathtools,graphicx,dcolumn}
\usepackage{enumerate,aas_macros}
\DeclareMathAlphabet{\mathpzc}{OT1}{pzc}{m}{it}

\newcommand{\sayy}[1]{`#1'}
\providecommand{\href}[2]{#2}

\makeatletter
\newcommand*{\rom}[1]{\expandafter\@slowromancap\romannumeral #1@}
\makeatother

\def\be{\begin{equation}}
\def\ee{\end{equation}}
\def\bea{\begin{eqnarray}}
\def\eea{\end{eqnarray}}

\def\sig{\sigma}

\def\la{\langle}
\def\ra{\rangle}
\def\Eu{ \mathfrak{H} }

\def\obs{\mathcal{O}}

\newcommand{\lcdm}{$\Lambda$CDM}

\begin{document}
\title{A prediction for anisotropies in the nearby Hubble flow} 
\author[a]{Asta~Heinesen,} 
\author[b]{Hayley J. Macpherson} 

\emailAdd{asta.heinesen@ens--lyon.fr} 
\emailAdd{h.macpherson@damtp.cam.ac.uk}

\affiliation[a]{Univ Lyon, Ens de Lyon, Univ Lyon1, CNRS, Centre de Recherche Astrophysique de Lyon UMR5574, F--69007, Lyon, France} 
\affiliation[b]{Department of Applied Mathematics and Theoretical Physics, Cambridge CB3 0WA, UK}

\abstract{ 
We assess the dominant low-redshift anisotropic signatures in the distance-redshift relation and redshift drift signals. We adopt general-relativistic irrotational dust models allowing for gravitational radiation---the \sayy{quiet universe models}---which are {extensions} of the silent universe models. Using cosmological simulations evolved with numerical relativity, we confirm {that} the quiet universe {model is} a good description on scales {larger than those of collapsing structures.} 
With this result, we reduce the number of degrees of freedom in the fully general luminosity distance and redshift drift cosmographies by a factor of $\sim 2$ and $\sim 2.5$, respectively, for the most simplified case. 
We predict a dominant dipolar signature in the distance-redshift relation for low-redshift data, with direction along the gradient of the large-scale density field. Further, we predict a dominant quadrupole in the anisotropy of the redshift drift signal, which is sourced by the electric Weyl curvature tensor. The signals we predict in this work should be tested with present and near-future cosmological surveys.

}

\maketitle

\section{Introduction} 
Cosmological data is most often interpreted within the Friedmann-Lema\^{\i}tre-Robertson-Walker (FLRW) metric models, which are characterised by their maximal {number of} symmetries  
over preferred spatial sections of the space-time. These models form the basis of the current standard cosmological model---the $\Lambda$ Cold Dark Matter (\lcdm) model. 
Low-redshift analyses commonly adopt FLRW cosmography: a formulation of 
nearby observables which explicitly depends on an FLRW geometry but is 
independent of the field equations that govern the expansion {of space} 
\cite{Visser:2003vq}. 
However, the low-redshift Universe is known to contain regional anisotropies from local {density contrasts and} matter flows. 
In order to take these into account in cosmological data analysis, one must go beyond the FLRW geometric ansatz. 
One method is to consider perturbations around a background FLRW metric, 
however, we might instead want to remain agnostic towards the particularities of the underlying (background) metric of the Universe. 
For this purpose we can use \emph{general} cosmography, where the form of the metric is left unspecified  \cite{KristianSachs,ELLIS1985315,EllisMacCallum,Clarkson:2011br,Umeh:2013UCT}. 

Exact multipole decompositions have been formulated for the general cosmographic expressions for luminosity distance 
\citep[up to third order in redshift;][]{Heinesen:2020bej} and redshift drift \citep[up to first order in redshift;][]{Heinesen:2021qnl}.  
The advantage of these formalisms is that they allow for model-independent data analysis of standardisable objects and redshift drift signals, 
and {for} inferring expansion and curvature invariants that describe our cosmic vicinity, without imposing {metric} symmetries {or constraints on the cosmological field equations}. The disadvantage is the large number of degrees of freedom (DOFs) that are {involved} when considering a fully general space-time description. 

In this paper, we consider a broad class of physical universe models which significantly reduce the number of DOFs 
characterising the cosmographies \cite{Heinesen:2020bej,Heinesen:2021qnl}, while still  being free of metric symmetries. 
Specifically, we consider an extension to the silent universe models \cite{Bruni:1994nf,vanElst:1996zs} considered in {\cite{Maartens:1996uv,Sopuerta:1998rt}}, which we denote\footnote{{The term \sayy{quiet universe} {was} used in \cite{Mutoh:1996ft} {for} silent universe models perturbed {with a small} magnetic Weyl curvature contribution. 
In this paper, we use the name \sayy{quiet universe} {to denote the extension} 
{which allows} for a magnetic Weyl curvature component that is not necessarily small, but constrained to have zero divergence.}} 
the {\sayy{quiet universe models}: 
irrotational dust space-times with vanishing divergence of the magnetic part of the Weyl tensor. 
We present the cosmographic relations for luminosity distance and redshift drift within {these models,} 
and confirm} the applicability of this class of models in describing a realistic cosmological setting by assessing the key constraints of such models in fully relativistic simulations.

\vspace{3pt} 
\noindent
\underbar{Notation and conventions:}
We use units in which the speed of light $c=1$ 
and the Einstein gravitational constant is 
$\kappa = 8 \pi G$, where $G$ is the Newtonian constant of gravitation. 
Greek letters $\mu, \nu, \ldots$ label space-time
indices in a general basis and repeated indices imply Einstein summation. 
The signature of the space-time metric $g_{\mu \nu}$ is $(- + + +)$ and  
$\nabla_\mu$ is the Levi-Civita connection. 
The permutation tensor $\epsilon_{\alpha\beta\gamma\delta}$ is defined as being equal to $\sqrt{-g}$ for even and $-\sqrt{-g}$ for odd
permutations of 0123, where $g$ is the determinant of the 
spacetime metric.
Round brackets, $(\, )$, containing indices denote symmetrisation in the involved indices and square brackets, $[\, ]$, denote anti-symmetrisation. 
We occasionally use bold notation, $\bm V$, for the basis-free representation of vectors $V^\mu$. 

\section{{The quiet universe models}} 
\label{sec:theory} 
Following {\cite{Maartens:1996uv,Sopuerta:1998rt}}, we consider a general-relativistic space-time where the energy-momentum content is well described by an irrotational dust source and  
the divergence of the magnetic part of the Weyl tensor is zero.
These constraints imply 
\begin{align}
    T_{\mu \nu} = \rho \, u_\mu u_\nu \, ,  \qquad &   \omega_{\mu \nu} \equiv h^\beta_{\, [ \mu} h^\alpha_{\, \nu ]} \nabla_{\alpha} u_{\beta} = 0 \, , \label{def:dust} \\ 
    D^\nu H_{\mu \nu} &= 0 \, , \label{def:nablaH}
\end{align}
where $T_{\mu \nu}$ is the energy-momentum tensor, $\bm u$ is the 4--velocity field of the congruence constituting the matter frame, $\omega_{\mu \nu}$ is the vorticity, and $H_{\alpha\beta} \equiv
-\frac{1}{2}\epsilon_{\rho\sigma\gamma\delta}
C_{\mu\nu}{}^{\gamma\delta}u^{\rho}h_{\alpha}^{\, \sigma}
u^{\mu}h_{\beta}^{\, \nu}$ is the magnetic part of the Weyl tensor in the matter frame. 
The magnetic and electric parts of the Weyl tensor together fully specify  
the Weyl curvature tensor $C_{\mu\nu \gamma\delta}$ \citep[see][for a review of the decomposition of the Weyl curvature tensor into electric and magnetic parts]{Maartens:1996ch}. 
The projector $h^\mu_{\, \nu} \equiv g^\mu_{\, \nu} + u^\mu u_{ \nu}$  
is the spatial metric on hypersurfaces orthogonal to the flow of $\bm u$. 
With the vanishing of vorticity, the kinematic decomposition of the matter frame yields 
\begin{align} \label{def:expu}
        \nabla_{\nu}u_\mu  &= \frac{1}{3}\theta\, h_{\mu \nu }+\sig_{\mu \nu} \, , \qquad \theta \equiv \nabla_{\mu}u^{\mu} \, , \qquad \sig_{\mu \nu} \equiv h^\beta_{\, ( \mu} h^\alpha_{\, \nu ) } \nabla_{\alpha} u_{\beta} - \frac{1}{3} \theta\, h_{\mu \nu} \, , 
\end{align} 
where $\theta$ is the volume expansion rate and $\sig_{\mu \nu}$ is the volume shear rate describing the anisotropic deformation {of the matter frame}. 

{We denote the class of general-relativistic space-times satisfying \eqref{def:dust} and \eqref{def:nablaH} \sayy{quiet universe models}.}   
Contrary to the class of silent universe models  \cite{Bruni:1994nf,vanElst:1996zs} in which $H_{\mu \nu}=0$, the weaker condition \eqref{def:nablaH} allows for gravitational radiation \cite{Matarrese:1993zf,Dunsby:1997fyr}.   
It might at first glance seem reasonable to neglect gravitational radiation for formulating a leading order cosmological model for approximating the late epoch Universe. However, 
the silent universe approximation is subject to a linearisation instability \citep{vanElst:1996zs} and is therefore not suitable {for describing} the non-linear regime of density contrasts.  
Furthermore, small values of $H_{\mu \nu}$ can allow for arbitrary ratios of shear eigenvalues \cite{Matarrese:1993zf}, 
breaking the axisymmetric expansion of fluid elements 
in the silent universe models \cite{vanElst:1996zs,Bolejko:2017wfy}. 
For this reason, we consider the broader class of universe models, where $H_{\mu \nu}$ is not constrained to be zero, and its curl can be non-zero.
The divergence-free condition (\ref{def:nablaH}) is stable under the exact evolution equations for an irrotational dust universe \cite{Maartens:1996uv} {provided that a chain of integrability constraints are satisfied \cite{Sopuerta:1998rt}.} 
{As remarked in \cite{Sopuerta:1998rt}, these integrability constraints are in general not satisfied in non-linear theory. {The} divergence-free condition \eqref{def:nablaH} holds in first order Lagrangian perturbation theory \cite{AlRoumi:2017gri}, which includes non-linear effects as compared to the standard perturbation theory approach. We might thus expect the quiet universe assumption to hold in the linear and slightly non-linear regime of density contrasts.} 

From the spatial parts of the Ricci identities, 
the kinematic variables of $\bm u$ satisfy the following constraints \cite{Elst:1996} 
\begin{align}
    D^\nu \sigma_{\mu \nu} &= \frac{2}{3} D_\mu \theta  \, , \label{brc1H} \\ 
     H_{\mu \nu} &= - h^{\rho}\!_{(\mu}\,h^{\sig}\!_{\nu)}
    \,\epsilon_{\rho\tau\kappa\lambda}\,(\nabla^{\tau}
    \sig^{\kappa}\!_{\sig})\,u^{\lambda} \, .
\label{csilhconstrH}  
\end{align} 
From the Bianchi identities, the electric part of the Weyl tensor, $$E_{\alpha\beta} \equiv C_{\mu\nu\rho\sigma}
u^{\mu}h_{\alpha}{}^{\nu}u^{\rho}h_{\beta}{}^{\sigma},$$ satisfies 
\begin{align}
    D^\nu E_{\mu \nu} &= \frac{1}{3} \kappa D_\mu \rho  - \epsilon_{\mu}^{\; \; \nu \rho \sigma} \sigma_{\nu \tau} H^{\tau}_{\, \rho} u_\sigma  \, , \label{brc4H} \\
    h^{\rho}\!_{(\mu}\,h^{\sig}\!_{\nu)}
    \,\epsilon_{\rho\tau\kappa\lambda}\,(\nabla^{\tau}  E^{\kappa}\!_{\sig})\,u^{\lambda} \!  &= \!   \nabla_\rho (u^\rho H_{\mu \nu} ) \!  - \!  3 \sigma^{\rho}_{\, \la \mu} H_{\nu \ra \rho}   \, , \label{csilhdotH} \\  
    \epsilon^{\mu \nu \rho \sigma} \sigma_{\nu \tau} E^{\tau}_{\, \rho} u_\sigma &= 0       \,  . \label{brc5H} 
\end{align} 
From \eqref{csilhdotH}, we see that a non-zero $H_{\mu \nu}$ allows for a non-zero curl of $E_{\mu \nu}$. The non-zero curls of the electric and magnetic Weyl tensors can be viewed as covariant requirements for gravitational wave propagation \cite{Maartens:1996ch}. 
We also see from (\ref{csilhconstrH}) that the curl of the shear tensor is non-zero in general and fully specifies $H_{\mu \nu}$. The magnetic Weyl tensor in turn enters in (\ref{brc4H}), where the right-most term thus represents the failure of the eigenbases of the shear tensor and its curl to be aligned. 
Equation (\ref{brc5H}) further implies that the eigenbasis of $E_{\mu \nu}$ and the eigenbasis of $\sigma_{\mu \nu}$ are aligned\footnote{This property is preserved from the silent universe model approximation. See \cite{Barnes_1989,vanElst:1996zs} for details.}. 
Invoking the evolution equations for shear and the electric Weyl tensor, we find the stronger condition: 
{the eigenbases of the shear tensor, the electric Weyl tensor, the curl of the magnetic Weyl tensor}, and all of their time derivatives are aligned \cite{Maartens:1996uv,Sopuerta:1998rt}. 
{The form of the constraint equations \eqref{brc1H}--\eqref{brc5H} remain unchanged with the introduction of a cosmological constant, as do the evolution equations for the shear and the electric Weyl components \cite{Elst:1996}. 
The properties of the quiet universe models described here thus extend to space-times that include a cosmological constant.

\section{Testing assumptions with relativistic simulations}

To examine the application of the quiet universe models to a realistic space-time, we will use cosmological simulations evolved {with} numerical relativity (NR) {using} realistic initial data. 
We describe the software we use in Section~\ref{sec:software}, our initial data in Section~\ref{sec:ics}, and our calculations assessing the {validity of the} quiet {and silent} universe approximations {in our simulations} in Section~\ref{sec:resultssim}. 

\subsection{Software}\label{sec:software}

We use the Einstein Toolkit\footnote{\url{https://einsteintoolkit.org}} \citep[ET;][]{Loffler:2012,Zilhao:2013}, a free, open source NR code based on the Cactus\footnote{\url{https://cactuscode.org}} infrastructure. The ET has been proven to be a useful tool for cosmological simulations {of large-scale structure formation,} without the need to define a fictitious background space-time \citep{Bentivegna:2017a,Macpherson:2017,Macpherson:2019a,Wang:2018}. 

The Einstein equations are evolved using the well-established BSSNOK formalism of NR \citep{Nakamura:1987,Baumgarte:1999,Shibata:1995}. {We} evolve the space-time metric using the \texttt{McLachlan} codes \citep{mclachlan}, the hydrodynamics using \texttt{GRHydro} \citep{grhydro} with a near-dust 
equation of state\footnote{There is a small {amount of} pressure {in the simulations,} however, the barotropic equation of state is chosen such that the pressure remains negligible. {This setup} has proven to be sufficient to match evolution of a dust FLRW model \citep{Macpherson:2017}.}, and set initial data set using \texttt{FLRWSolver}\footnote{\url{https://github.com/hayleyjm/FLRWSolver_public}} \citep{Macpherson:2017}. {We use a harmonic-type evolution of the lapse function and set the shift to zero throughout \citep[see][for details of the gauge we use]{Macpherson:2019a}.}
The cosmological constant is set to zero in our simulations, 
since {the codes we use were} 
originally intended for relativistic systems on small scales where dark energy can safely be ignored.
Since our simulations are matter dominated, we will have naturally higher density contrasts on the ``present-epoch''\footnote{See Section~\ref{sec:resultssim} for the definition of the ``present-epoch'' hypersurface in our simulations.} hypersurface with respect to a \lcdm\ model universe. 
{We expect our results to be valid for \lcdm\ cosmology on scales with comparable density contrasts. } 

\subsection{Initial data}
\label{sec:ics}

\texttt{FLRWSolver} specifies linear perturbations atop a flat FLRW background with dust source, drawn from a user-provided matter power spectrum at the chosen initial redshift \citep[see][for more details]{Macpherson:2019a}. The section of the power spectrum that is used depends on the physical size of the box and the grid resolution. 
In this paper, we generate the matter power spectrum of perturbations using the CLASS\footnote{\url{http://class-code.net}} code at our initial redshift $z_{\rm ini}=1000$.}

For the purpose of examining {general} cosmographic relations 
with the observed redshift as a parameter along photon null lines, we must cut out small-scale collapsing structures from our simulations\footnote{See \cite{Macpherson:2021a} for a discussion on smoothing scale in relation to cosmography.}. 

In \cite{Macpherson:2021a}, we studied the anisotropic signals in cosmological parameters in the general {luminosity distance} cosmography. {In that work,} 
we used simulations with individual grid cells with size 100--200 $h^{-1}$ Mpc in order to \emph{strictly} exclude any structure beneath these scales. 
In this work, we are interested in assessing the applicability of the quiet universe assumption in Section~\ref{sec:theory}, which involves evaluation of the shear and Weyl tensors.  
{Due to the under-sampling of structure,} we find that the numerical precision of the simulations we used in \cite{Macpherson:2021a} is not sufficient for calculations of the electric Weyl tensor, which has small numerical values and therefore is dominated by finite-difference and round-off errors. 
Therefore, in this work, we {mitigate this issue by ensuring} that the smallest-scale modes are sampled by at least 10 grid cells in the initial data for all simulations.

{In order to} perform numerical convergence studies and quantify errors on our results, we perform the same simulation at three resolutions $N=64, 128$, and 256, where $N^3$ is the total number of grid cells. 
The smallest-scale modes are {therefore} sampled by 10, 20, and 40 grid cells for the $N=64,128,$ and 256 resolution simulations{, respectively}. 
Excluding {modes beneath these scales} 
requires making a cut to the initial power spectrum, {i.e. setting}  
$P(k>k_{\rm cut})=0$, where $k_{\rm cut}=2\pi/\lambda_{\rm cut}$ and $\lambda_{\rm cut}$ is the minimum wavelength sampled.  
We choose $\lambda_{\rm cut}=200 h^{-1}$ Mpc {for our large-scale cosmological simulation}, 
{implying} that the 
grid spacing for $N=64$ {must be} $\Delta x_{64} = 20 h^{-1}$ Mpc, and our box size is thus $L=1280 h^{-1}$ Mpc for all resolutions. Here, we have defined {the Hubble constant at redshift zero to be} $H_0=100 h$ km/s/Mpc with $h=0.7$ to define the 
length scales of the simulation only, and no global FLRW Hubble expansion is enforced during the simulation. 
We note that while we cut {out} all modes below {the} $200 h^{-1}$ Mpc {scale} in the initial data, this does not in general prevent smaller-scale structures from forming later in the simulation. {However,} we have found that cutting modes at this scale still results in a quite smooth model universe at redshift zero. 

We must ensure {that} derivatives are consistent between simulations to perform direct numerical convergence studies at individual grid points. We therefore generate the initial density field for the 
$N=64$ simulation 
{and} interpolate this field to $N=128$ and 256 before solving the {general-relativistic constraint equations} to linear order. This procedure fully specifies the initial Cauchy surface of our simulations. 

We wish to test the applicability of {the quiet universe model} in the {large-scale} simulations described above, {in order to assess the applicability of this model for describing large-scale cosmography of observables.}  
{We are also interested in testing} the applicability of {this model} 
in the nonlinear regime of structure growth. To this end, we also analyse a simulation which samples smaller scales than those described above.  
Specifically, we perform {one} simulation with $N=256$ and 
{$L=1592\,h^{-1}$Mpc,} 
maintaining the requirement that all modes be sampled with at least 10 grid cells in the initial data{---  
sampling} modes down to 
{$62\,h^{-1}$Mpc} 
scales. Otherwise, the initial data for this simulation is generated in the same way as outlined above.

{While} the initial data is assumed as linear perturbations around an FLRW background, we stress that the simulation itself is not explicitly constrained to follow an FLRW evolution. 
However, in terms of global averages we find excellent agreement with the Einstein-de Sitter (EdS) model \citep[see also][]{Macpherson:2019a}. 
Specifically, for all 
simulations 
we find the globally-averaged, present-epoch cosmological parameters \citep[see][for definitions of these]{Macpherson:2019a} to be consistent with the EdS values $\Omega_m=1, \Omega_R=0,$ and $\Omega_Q=0$ {to within the numerical errors of the simulation}. {We find} that the globally-averaged Hubble parameter $\mathcal{H}_{\rm all}\equiv \langle \theta \rangle_{\rm all} /3${---where $\langle\rangle_{\rm all}$ indicates an average over the entire {present epoch} simulation domain---}is {also} consistent with the EdS value of $H_0\approx 45$ km/s/Mpc to 
{within numerical errors.}

\subsection{{Testing approximations in the simulations}} 
\label{sec:resultssim}
We evolve the Einstein equations from the initial Cauchy surface until a present epoch hypersurface, 
where {we define} the \sayy{present}  
as {the} hypersurface where average length scales have increased by a factor {$\sim (1+z_{\rm ini})$}  
relative to those on the initial surface.

The vorticity-free condition \eqref{def:dust} and the magnetic Weyl curvature condition \eqref{def:nablaH} are not satisfied identically, but we expect them to remain approximately satisfied for the large scales {we consider.}
{The 
constraint equations \eqref{brc1H}--\eqref{brc5H} are of particular interest for simplifying the cosmographic expressions {in} 
\cite{Heinesen:2020bej,Heinesen:2021qnl}.}
Here we quantify in detail the applicability of {these} constraints in {our  
large-scale} simulations. 
We also examine some additional properties, which do not immediately follow from \eqref{def:dust} and \eqref{def:nablaH}, but which might be useful {for {further} simplifying the} {cosmography}.

\begin{figure*}
    \centering
    \includegraphics[width=\textwidth]{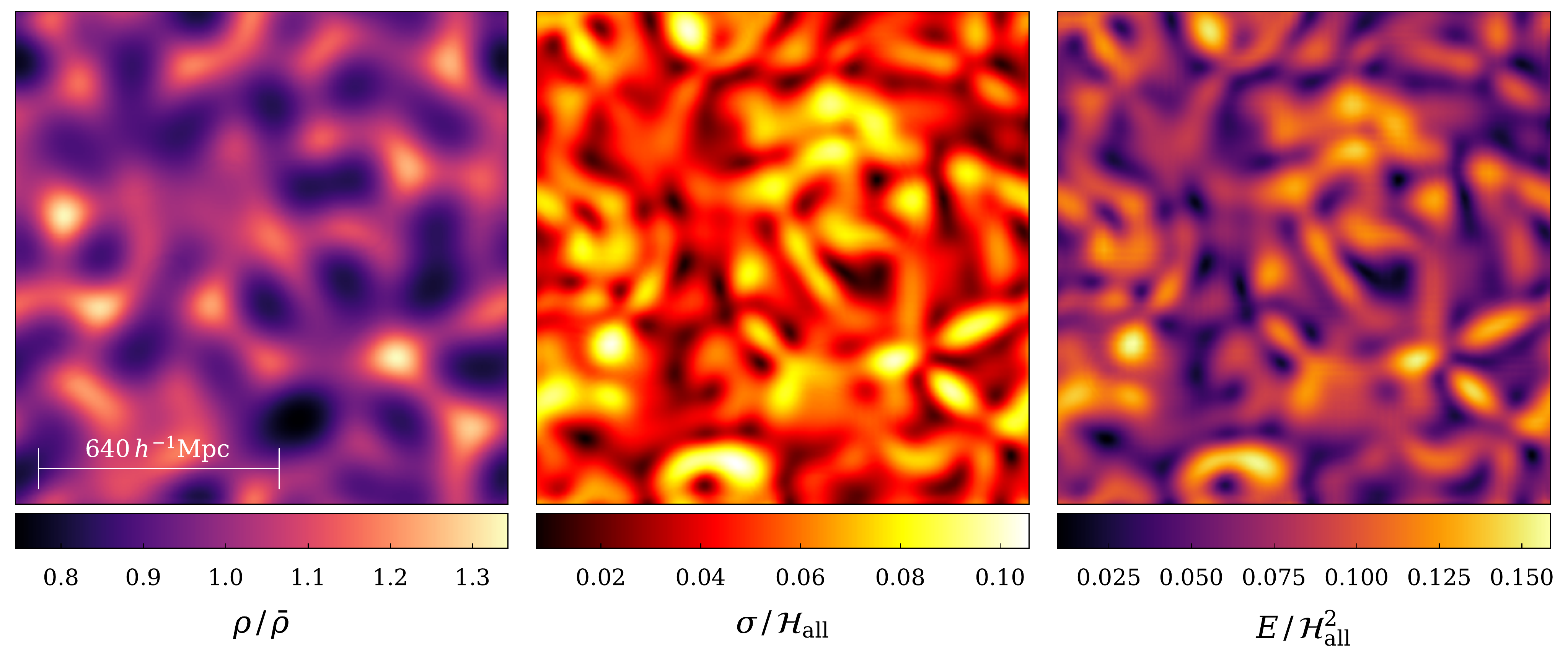}
    \caption{2--dimensional slices in the $N=256$ simulation with all structure beneath 200$h^{-1}$ Mpc cut out of the initial data. Left panel shows the density relative to the mean over the whole box, middle shows the shear scalar normalised by the globally-averaged Hubble rate, and the right panel shows the Weyl scalar also normalised by the Hubble rate.}
    \label{fig:200Mpc_2Dpanels}
\end{figure*} 
Figure~\ref{fig:200Mpc_2Dpanels} shows 2--dimensional slices of the density relative to the global average, $\rho/\bar\rho$, and the shear and electric Weyl {scalar} fields, 
\begin{align}\label{eq:Eandshear}
    \sigma^2 \equiv \frac{1}{2}\sigma^{\mu\nu} \sigma_{\mu\nu} \, , \qquad E^2 \equiv \frac{1}{2} E^{\mu\nu} E_{\mu\nu}, 
\end{align}
from left to right, respectively{ in the large-scale simulation}. {The shear and Weyl scalars are} normalised by $\mathcal{H}_{\rm all}$ such that they are dimensionless. 
Typical density contrasts of the {large-scale} simulation are {$\sigma_\delta\approx0.09$, and for the simulation sampling smaller scales, these are $\sigma_\delta\approx 3$.} 
{These values are} higher than what would be expected for a $\Lambda$CDM universe as seen on {similar} comoving scales, which 
can be explained by two main effects. Firstly, {our} simulations 
do not {have} 
a cosmological constant, which means that the focusing of structure is not counteracted by a negative pressure component. 
Secondly{, as remarked in Section~\ref{sec:ics},} while initial conditions are featureless below the comoving scale 200$h^{-1}$ Mpc, this does not prevent structure at smaller scales from forming later in the evolution. {In fact, features below the $\sim 200\,h^{-1}$Mpc scale are visible in Figure~\ref{fig:200Mpc_2Dpanels}.}

\subsubsection{Alignment of shear and electric Weyl eigenbases}
 
{We} will examine the alignment of the shear and electric Weyl tensor bases as prescribed by the relation (\ref{brc5H}). 
First, we compute the dimensionless commutation index  
\begin{eqnarray}
&& \mathcal{C} \equiv \frac{\mathcal{A}^{[\mu \nu]} \mathcal{A}_{[\mu \nu]}}{\mathcal{A}^{\mu \nu} \mathcal{A}_{\mu \nu}}  \, , \qquad  \mathcal{A}_{\mu \nu} \equiv \sigma_{\tau \mu} E^{\tau}_{\, \nu }         \,  , \label{brc5Htest} 
\end{eqnarray} 
which equals zero {\emph{only}} if (\ref{brc5H}) is satisfied, and equals one in the opposite extreme case: where anti-commutation of $\sigma_{\mu \nu}$ and $E_{\mu \nu}$ is satisfied. 

\begin{figure}
\label{fig:Cindex}
     \centering
     \begin{subfigure}[b]{0.49\columnwidth}
         \centering
         \includegraphics[width=1\columnwidth]{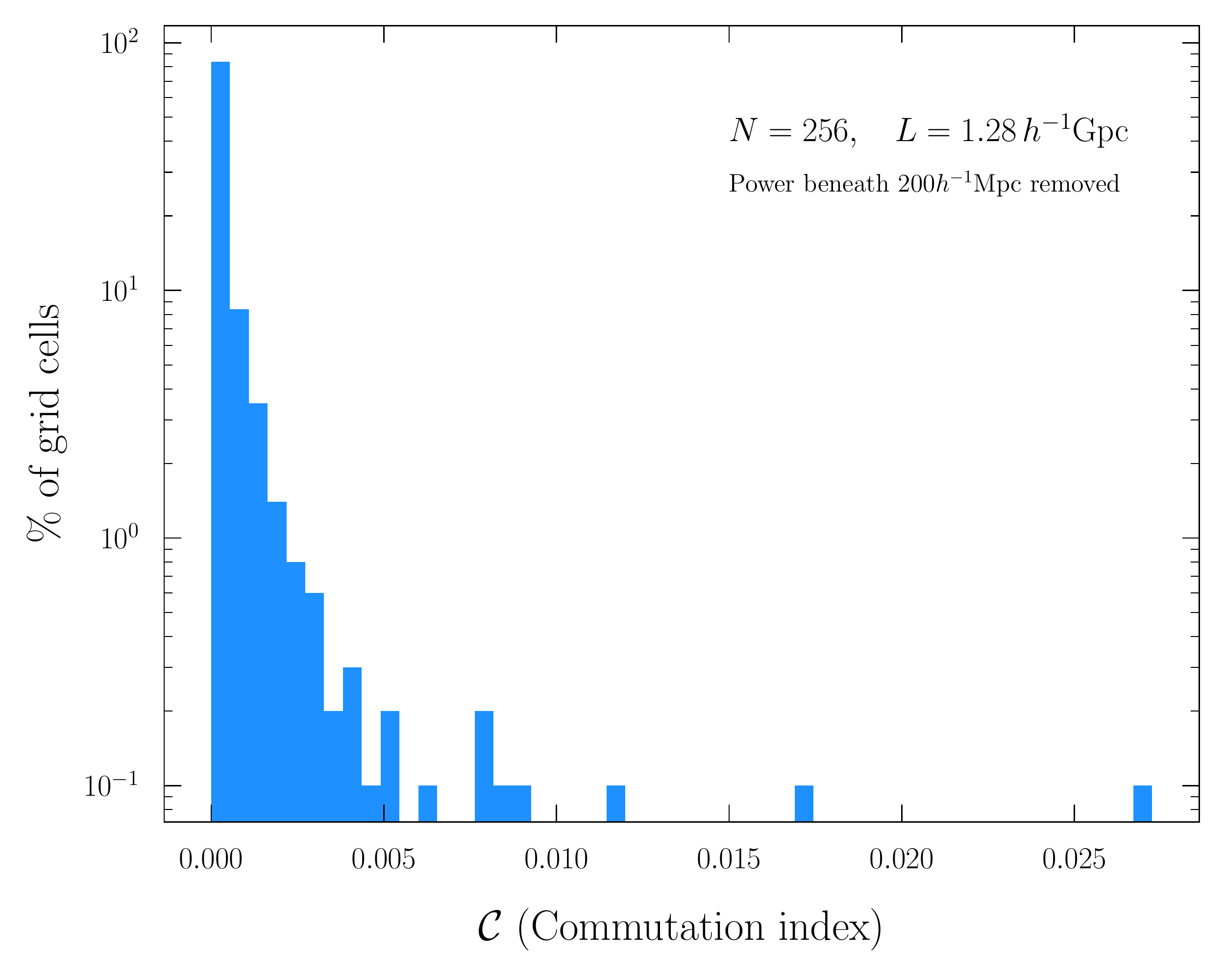}
         \caption{$200 h^{-1}$ Mpc cutoff in scale.}
         \label{fig:Cindex_200}
     \end{subfigure}
     \hfill
     \begin{subfigure}[b]{0.49\columnwidth}
         \centering
         \includegraphics[width=1\columnwidth]{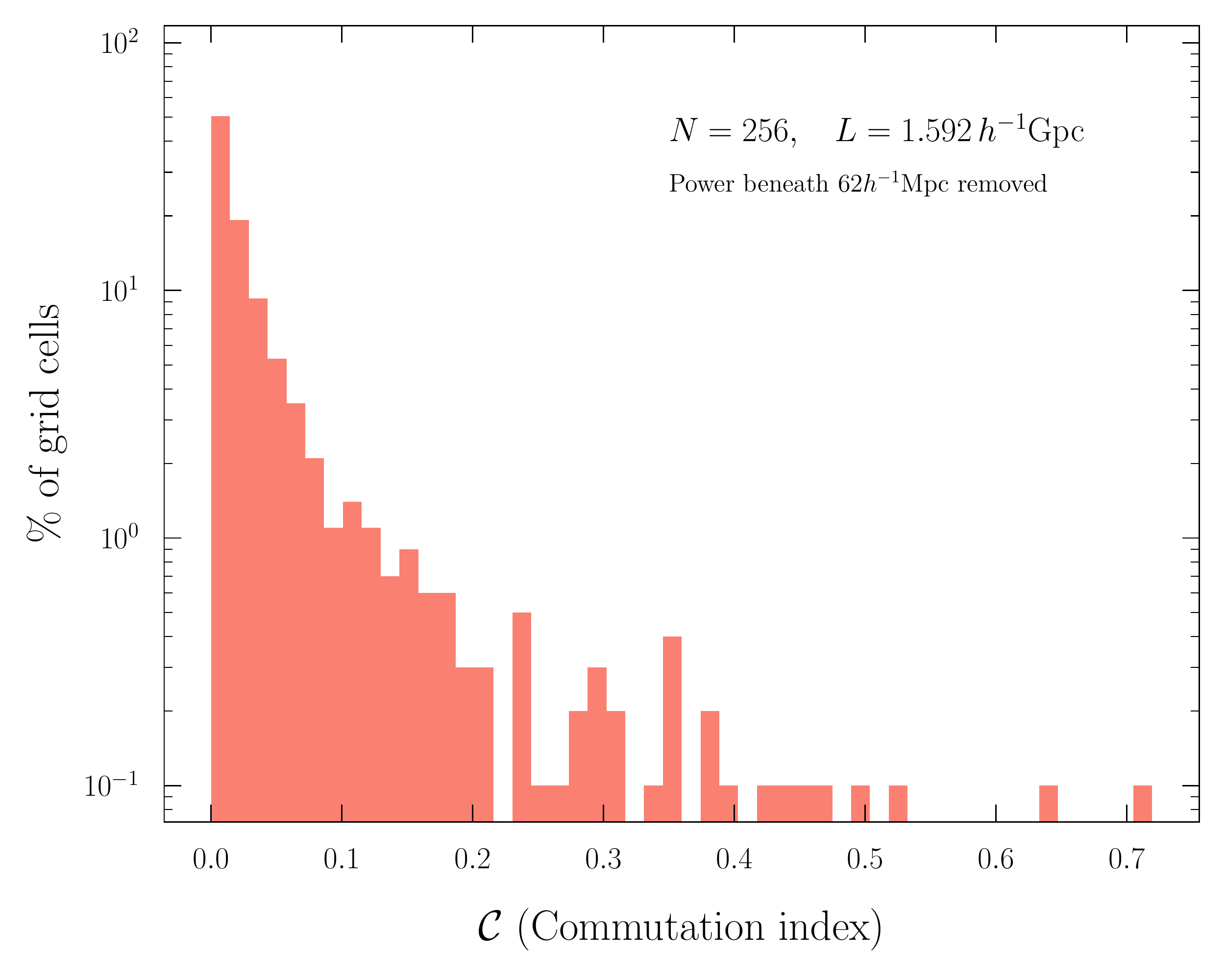}
         \caption{{$62 h^{-1}$ Mpc} cutoff in scale.}
         \label{fig:Cindex_40}
     \end{subfigure}
        \caption{The commutation index, $\mathcal{C}$, as show for 1000 grid points over the present epoch surface simulation domain. $\mathcal{C}=0$ indicate commutation of $E_{\mu \nu}$ and $\sigma_{\mu \nu}$ whereas $\mathcal{C}=1$ indicate anti-commutation.}
\end{figure} 
{Figure~\ref{fig:Cindex_200} shows the commutation index of the {large-scale} simulation at 1000 {evenly spaced} grid points, 
for which we see anti-commutation 
at the $< 0.005$ level for $\sim 99\%$ of the grid points. 
{Figure~\ref{fig:Cindex_40} shows the same commutation index calculated at 1000 evenly-space grid points in the simulation sampling down to 62$h^{-1}$Mpc in the initial data{, where the anti-commutation is at the $<0.1$ level for $\sim 91\%$ of the grid points.}

\begin{figure}
\label{fig:eig1}
     \centering
     \begin{subfigure}[b]{0.49\columnwidth}
         \centering
         \includegraphics[width=1\columnwidth]{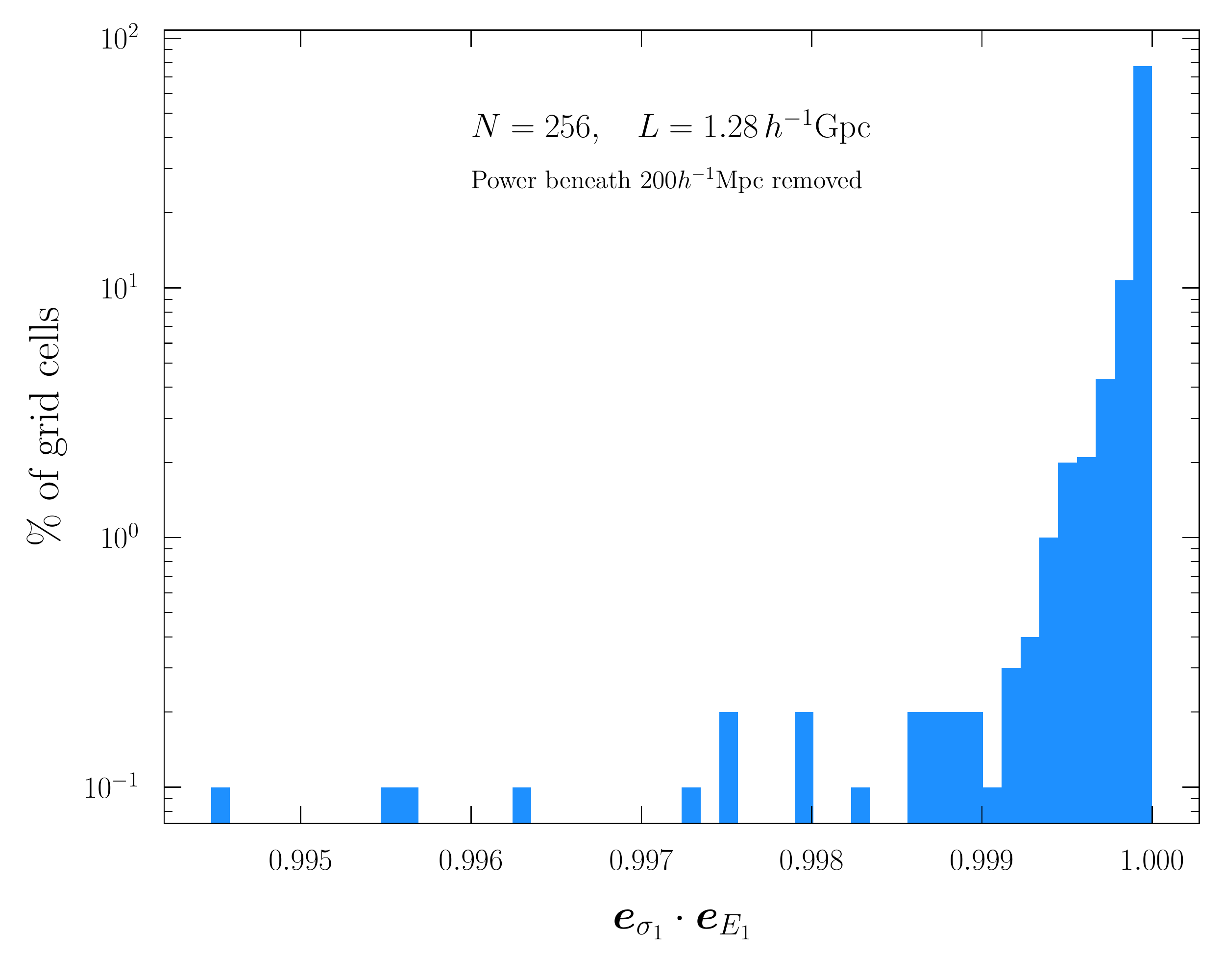}
         \caption{$200 h^{-1}$ Mpc cutoff in scale.}
         \label{fig:eig1_200}
     \end{subfigure}
     \hfill
     \begin{subfigure}[b]{0.49\columnwidth}
         \centering
         \includegraphics[width=1\columnwidth]{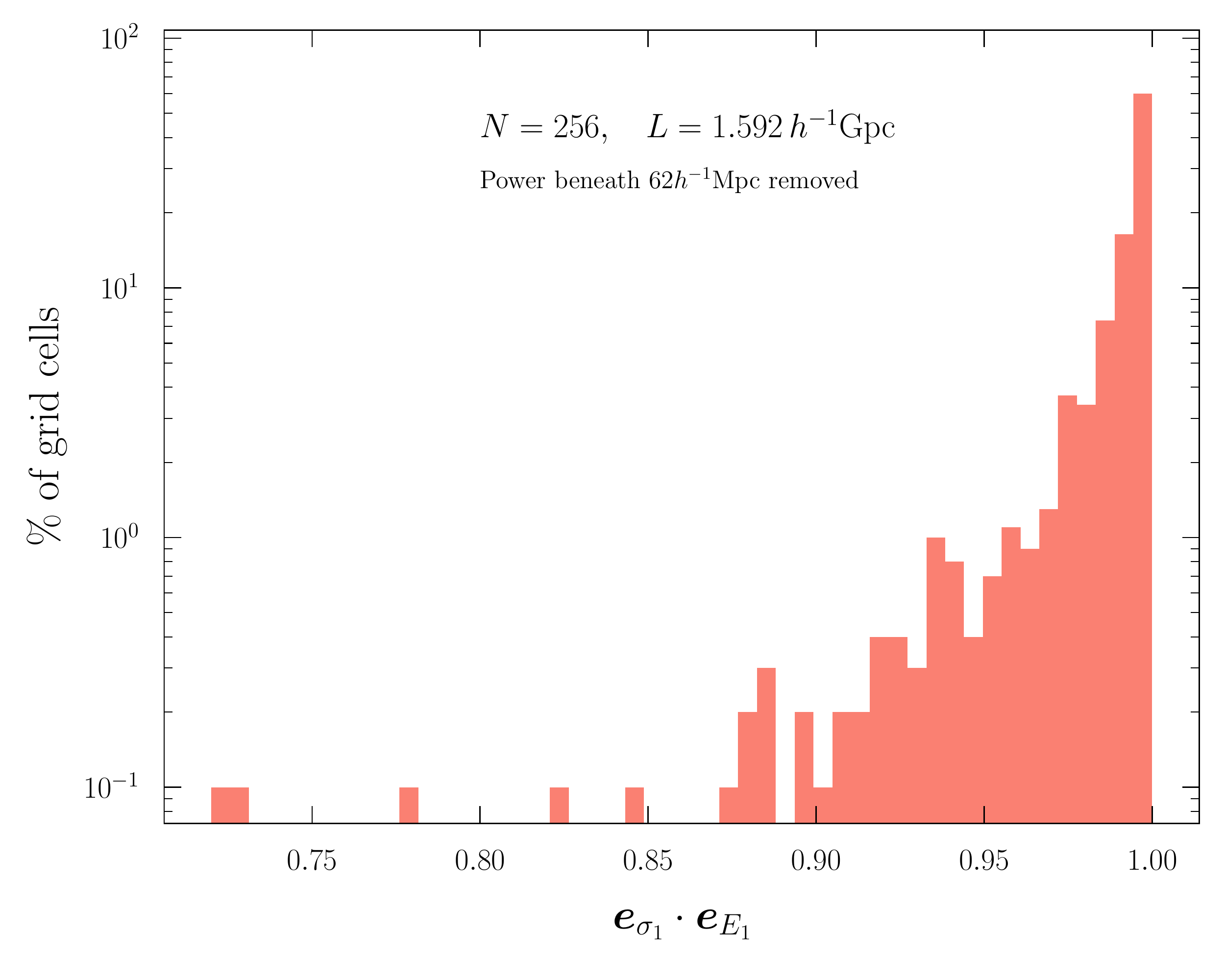}
         \caption{{$62 h^{-1}$} Mpc cutoff in scale.}
         \label{fig:eig1_40}
     \end{subfigure}
        \caption{{The dot product between the principal eigendirection of the shear tensor, {$\bm e_{\sigma_1}$}, and the nearest eigendirection of the electric Weyl tensor, {$\bm e_{E_1}$}, as show for 1000 grid points over the present epoch surface simulation domain. When $\bm e_{\sigma_1}\cdot \bm e_{E_1} = 1$, there is perfect alignment between the two eigendirections.}  }
\end{figure} 

We can also visualise the alignment property 
by computing the {eigenbases} of the shear tensor and electric Weyl tensor.  
{We} solve the eigenvalue problem for both tensors and calculate the dot {products of their eigendirections} at each grid cell. Figure~\ref{fig:eig1_200} shows the dot product of the principal eigendirection of the shear tensor, {$\bm e_{\sigma_1}$}, with the nearest eigendirection of the electric Weyl tensor, {$\bm e_{E_1}$}, {for the simulation sampling down to 200$h^{-1}$Mpc, } 
showing alignment to within {$0.5\%$} for $\sim 99.9\%$ of grid points.
Figure~\ref{fig:eig1_200} shows the same calculation for the simulation sampling down to {62$h^{-1}$Mpc} in the initial data, where $\sim 99\%$ of the grid points show alignment to within $10\%$. 

We conclude, based on these two measures, that alignment of the eigenbases of the shear and electric Weyl tensors is a good approximation within our simulations smoothing over $\sim 200 \,h^{-1}$ Mpc. 
For comparison, {in the simulation containing smaller-scale} structure, 
the commutation index is still skewed towards alignment of the shear and Weyl tensor, but {less} so than for the large scale simulation. 
The {weakening of alignment between shear and electric Weyl eigenvectors is expected as} collapsing structures are resolved (as is the case 
in {this} 
simulation):  
the irrotational requirement of the fluid breaks down 
and divergences of $H_{\mu \nu}$ might become important. 
We note however that this level of coarsegraining is not immediately suited for cosmography, since the collapsing regions {cause a change of sign of $\Eu$ and thus the cosmographic relation breaks down.} 
Some level of (implicit) coarsegraining above scales of collapsing regions is needed for observables to be single valued functions in redshift. On cosmological scales, where expansion is dominating over rotation DOFs, we 
expect the shear-electric Weyl alignment property to be a good approximation,
{which we have verified in our large-scale simulations.} 

}

\subsubsection{Applicability of the silent universe approximation} 
\label{sec:silent}
{ 
{We now examine the applicability of the silent universe models \cite{Bruni:1994nf,vanElst:1996zs} {in describing} our 
simulations. 
The silent universe models belong to the class of quiet universe models in Section~\ref{sec:theory}, and are further constrained by the condition $H_{\mu \nu} = 0$. } 
An important consequence of the silent universe approximation is that {the two non-principal eigenvalues} of the shear tensor, {$\sigma_2$ and $\sigma_3$,} are degenerate, such that {their ratio} $\sigma_3/\sigma_2 = 1$.

\begin{figure}[!h] 
\label{fig:sheareig}
     \centering
     \begin{subfigure}[b]{0.49\columnwidth}
         \centering
         \includegraphics[width=1\columnwidth]{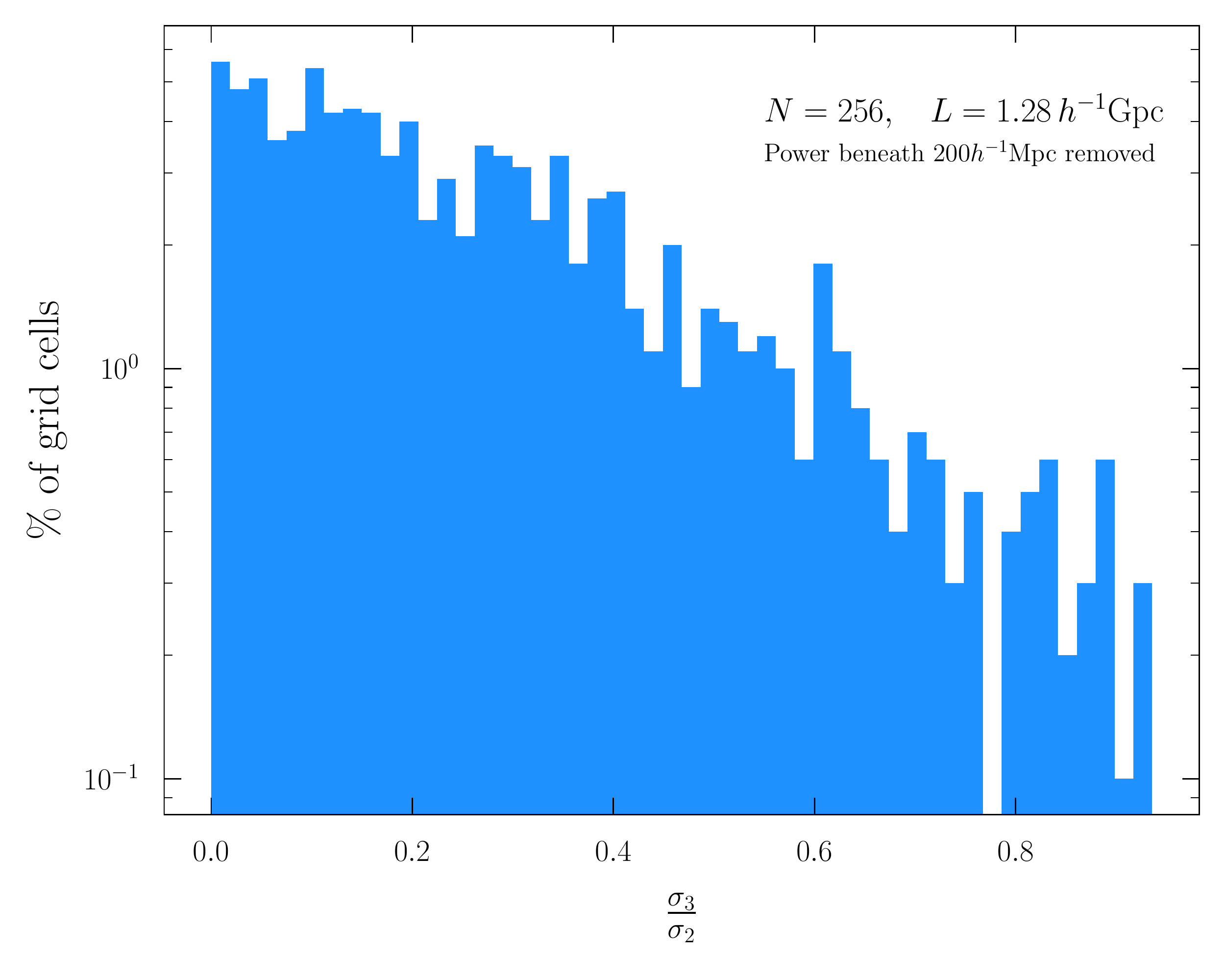}
         \caption{$200 h^{-1}$ Mpc cutoff in scale.}
         \label{fig:sheareig_200}
     \end{subfigure}
     \hfill
     \begin{subfigure}[b]{0.49\columnwidth}
         \centering
         \includegraphics[width=1\columnwidth]{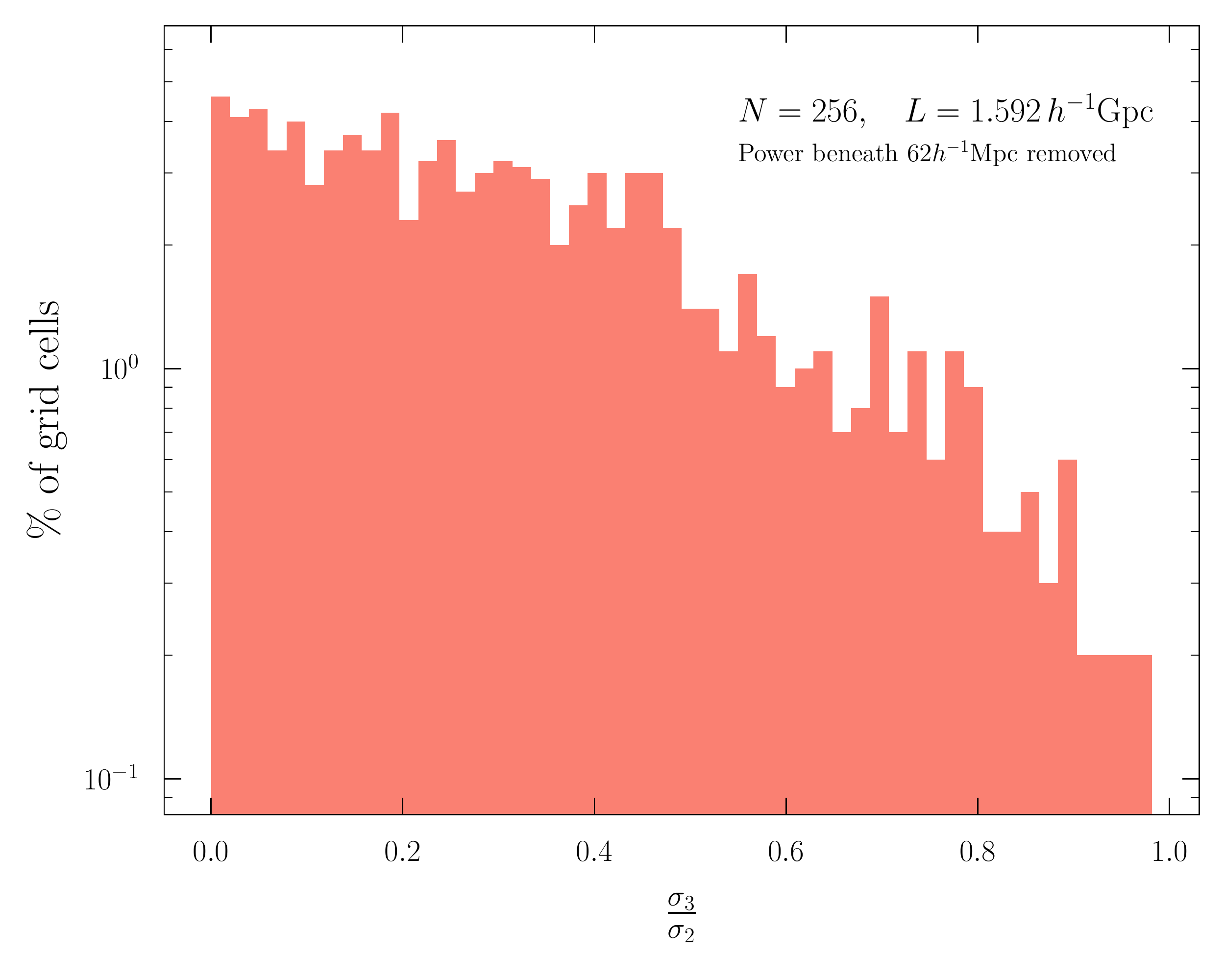}
         \caption{$62 h^{-1}$ Mpc cutoff in scale.}
         \label{fig:sheareig_40}
     \end{subfigure}
        \caption{{Ratio of the non-principal shear eigenvalues, $\sigma_2$ and $\sigma_3$,} as show for 1000 grid points over the present epoch surface simulation domain. {A value $\sigma_3/\sigma_2 = 1$ corresponds to degeneracy between the shear eigenvalues at the given grid point.}}
\end{figure} 
In {Figure~\ref{fig:sheareig_200}} we show the {ratio of the} non-principal eigenvalues of the shear {tensor} 
at 1000 {evenly spaced} grid points in the large-scale simulation. 
For most  
grid points, the ratio $\sigma_3/\sigma_2$ is closer to zero than to one, {with only $\sim4\%$ of the grid points having $\sigma_3/\sigma_2>0.9$.}  
{We thus} conclude that there is no (approximate) degeneracy between shear eigenvalues in the large-scale simulation. 
The simulation {with smaller-scale structure} shows a similar tendency, with no degeneracy between shear eigenvalues, {as shown} in Figure~\ref{fig:sheareig_40}.

We conclude that the silent universe approximation is broken, even for the large-scale simulation investigated here. 
{As is detailed in \cite{Matarrese:1993zf,vanElst:1996zs}, the silent universe models have a linearisation instability. However, it is not obvious that the silent approximation would be insufficient in our large-scale model  
universe, where the density field is close to the linear regime.}

Since the quiet universe provides a good description for our large-scale simulations, the  breakdown of the silent universe approximation must occur because of the breaking of the additional assumption of $H_{\mu \nu} = 0$. 
{The magnetic part of the Weyl tensor has no simple Newtonian counterpart \cite{Buchert:2012mb} and the limit of vanishing magnetic part of the Weyl tensor is {therefore} often considered Newtonian-like\footnote{The Newtonian limit of general relativity is non-trivial and has been argued to contain {magnetic} Weyl-type counterparts in general \cite{Bertschinger:1994nc,Kofman:1994pz,Ellis:1994md}.} {\cite{Maartens:1998ci,Ehlers:2009uv}}. Consequently, the failure of the silent universe approximation to apply could be assigned to purely general-relativistic effects. 
We note that even though the components of $H_{\mu \nu}$ are small, 
their impact on the breaking of the degeneracy of the shear eigenvalues {is} of order 1, as 
can be seen in {Figure~\ref{fig:sheareig_200} and Figure~\ref{fig:sheareig_40}}. 
It is an interesting result in its own right that the \sayy{weak field} (in the context of density contrasts) cosmological simulation considered here exhibits fundamentally general-relativistic properties.}  
{The breakdown of the silent universe models might have implications for the accuracy of Newtonian modelling of cosmological structure formation, cf. \cite{Mutoh:1996ft}.}

\subsubsection{Proportionality of the electric Weyl and shear tensors}
\label{sec:propEsigma}

{In the middle and right panels of Figure~\ref{fig:200Mpc_2Dpanels}, some level of correlation between $\sigma$ and $E$ is visible by eye. }
We can further examine the applicability of the proportionality law $E_{\mu \nu} \propto \sigma_{\mu \nu}$ which, on top of alignment of the eigenbases of $E_{\mu \nu}$ and $\sigma_{\mu \nu}$, also requires common proportionality between the eigenvalues, such that $\sigma_1/E_1 = \sigma_2/E_2 = \sigma_3/E_3$, where $\sigma_1$ is the principal eigenvalue of $\sigma_{\mu \nu}$, and $\sigma_2$ and $\sigma_3$ are the two remaining eigenvalues where $\sigma_3$ is the smallest in amplitude (analogous definitions hold for the eigenvalues $E_1$, $E_2$, and $E_3$ of $E_{\mu \nu}$). 

\begin{figure}[!h] 
\label{fig:propeig}
     \centering
     \begin{subfigure}[b]{0.49\columnwidth}
         \centering
         \includegraphics[width=1\columnwidth]{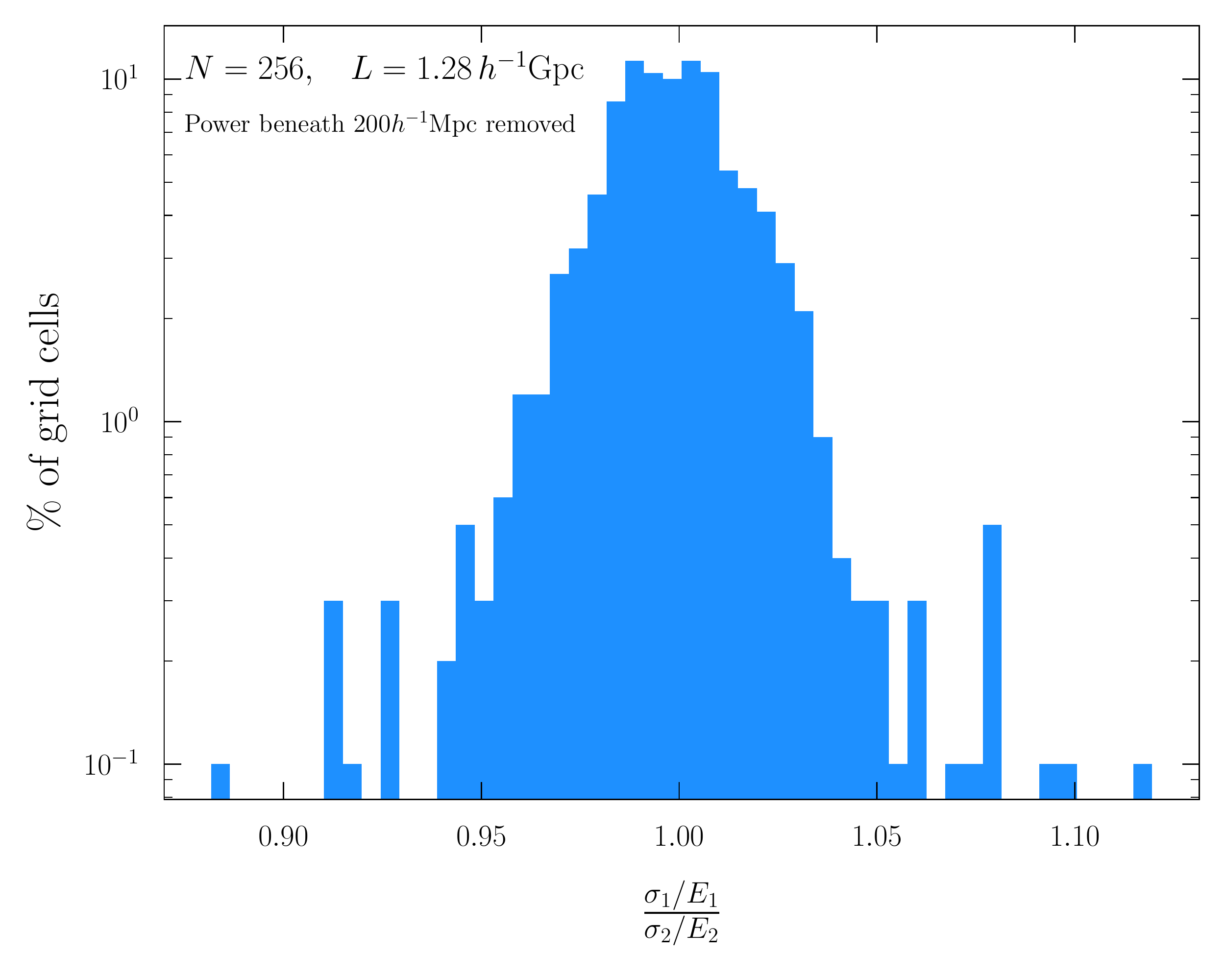}
         \caption{$200 h^{-1}$ Mpc cutoff in scale.}
         \label{fig:propeig_200}
     \end{subfigure}
     \hfill
     \begin{subfigure}[b]{0.49\columnwidth}
         \centering
         \includegraphics[width=1\columnwidth]{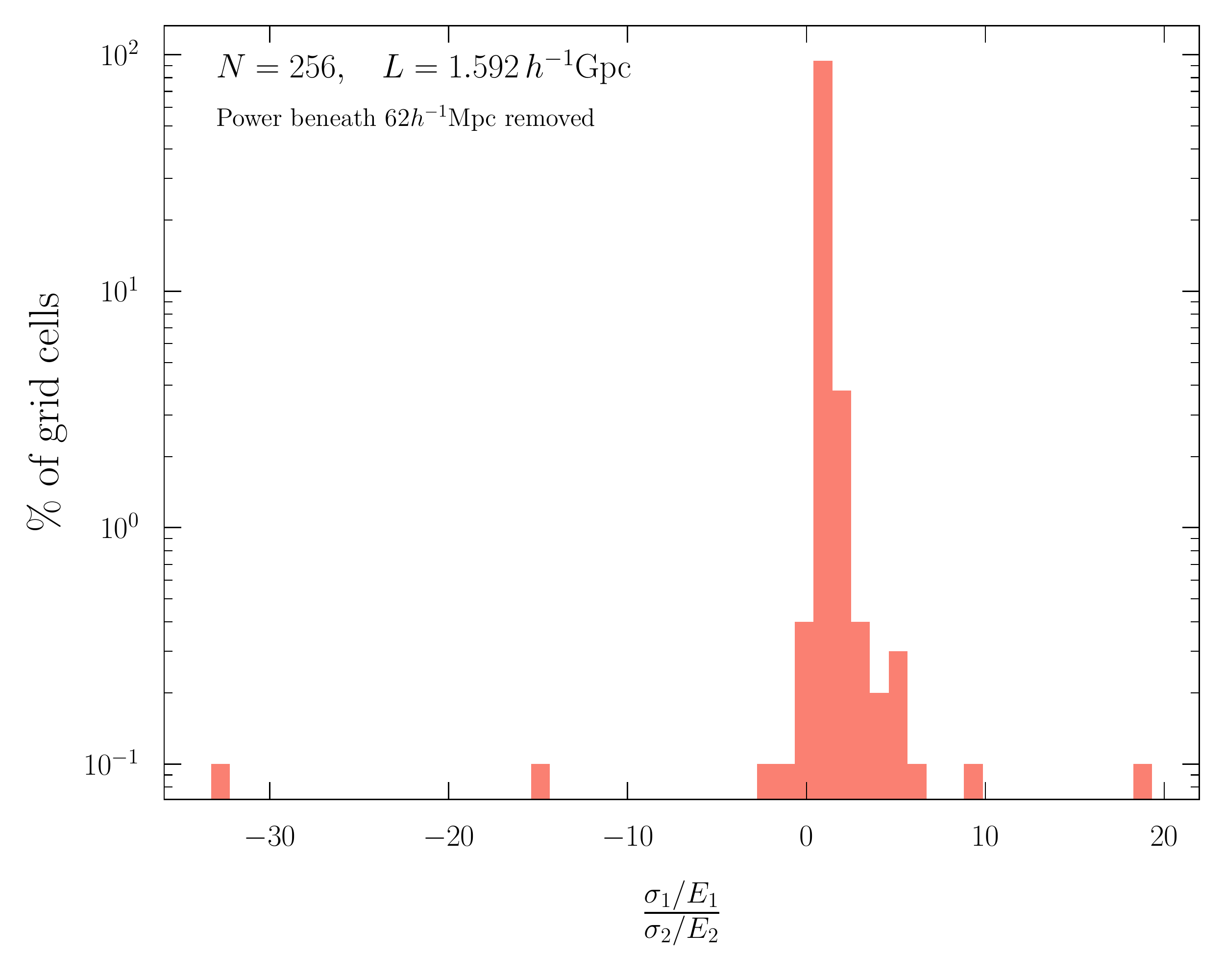}
         \caption{$62 h^{-1}$ Mpc cutoff in scale.}
         \label{fig:propeig_40}
     \end{subfigure}
        \caption{Ratio of the principal and second largest shear eigenvalues divided by the ratio of the corresponding electric Weyl curvature eigenvalues as shown for 1000 grid points over the present epoch surface simulation domain.}
\end{figure} 
{Figure~\ref{fig:propeig_200}} shows the ratio $(\sigma_1/E_1)/(\sigma_2/E_2)$ for {1000} grid points in the 
{large-scale} simulation. 
Departures from $(\sigma_1/E_1)/(\sigma_2/E_2) = 1$ are $\geq 5\%$ 
for $\sim 3$\% of the grid points. Departures are in general larger from $(\sigma_1/E_1)/(\sigma_3/E_3) = 1$ with departures of $\geq 50\%$ for $\sim 7\%$ of the grid points. However, the latter ratio involves the smaller eigenvalues, $\sigma_3$ and $E_3$, and is thus sensitive to small absolute fluctuations in either $\sigma_3$ or $E_3$. As a crude first order model assumption, we {can therefore} 
employ $E_{\mu \nu} \propto \sigma_{\mu \nu}$ for the majority of grid points. 
{The proportionality law between the shear and electric Weyl tensors is broken in the 
simulation {with smaller-scale structures}, as is shown in Figure~\ref{fig:propeig_40}, where $\sim60\%$ of the grid points have departures $\geq 10\%$ from $(\sigma_1/E_1)/(\sigma_2/E_2) = 1$.}   

Another way  
{we can test} the approximate proportionality between the shear tensor and the electric Weyl tensor is {to probe} the (anti-)alignment between $D_\mu \theta$ and $D_\mu \rho$. The proportionality $D_\mu \theta \propto D_\mu \rho$ is exact when the conditions (\ref{def:dust}) and (\ref{def:nablaH}) are fulfilled and when $E_{\mu \nu} = K \sigma_{\mu \nu}$, where $K$ is a constant in the fluid frame: $D_\mu K = 0$, and 
$\epsilon_{\mu}^{\; \; \nu \rho \sigma} \sigma_{\nu \tau} H^{\tau}_{\, \rho} u_\sigma = 0$ {(from (\ref{brc1H}) and (\ref{brc4H}))}.
{To} probe this alignment we calculate the {normalised} dot product between $D_\mu \theta$ and $D_\mu \rho$ in the fluid frame. 

\begin{figure}[!h] 
\label{fig:dotproducts}
     \centering
     \begin{subfigure}[b]{0.49\columnwidth}
         \centering
         \includegraphics[width=1\columnwidth]{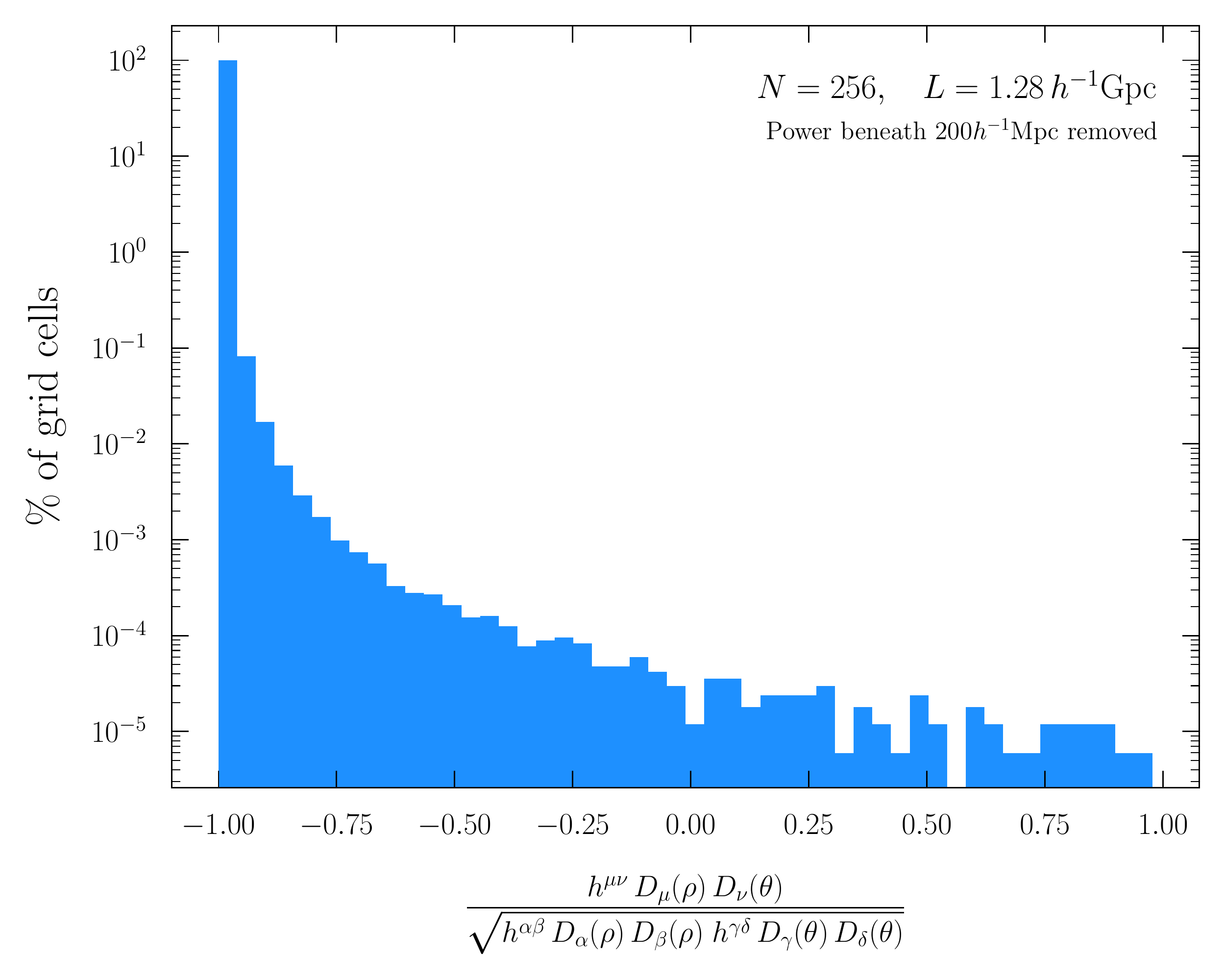}
         \caption{$200 h^{-1}$ Mpc cutoff in scale.}
         \label{fig:dots_200}
     \end{subfigure}
     \hfill
     \begin{subfigure}[b]{0.49\columnwidth}
         \centering
         \includegraphics[width=1\columnwidth]{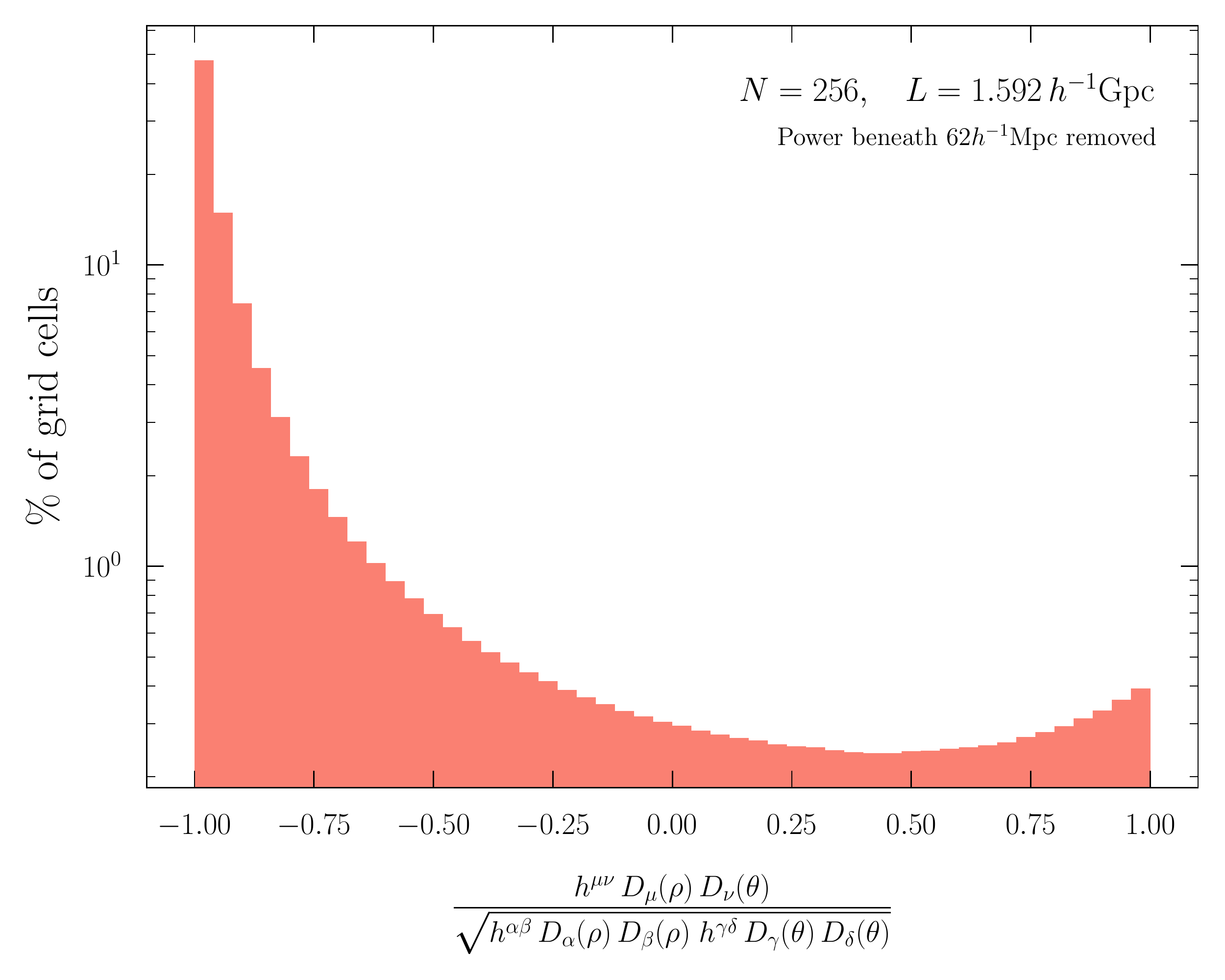}
         \caption{$62 h^{-1}$ Mpc cutoff in scale.}
         \label{fig:dots_40}
     \end{subfigure}
        \caption{Histogram of the alignment of $D_\mu \rho$ and $D_\mu \theta$ for all grid cells in the present-epoch surface simulation domain. 
        }
\end{figure}
{Figure~\ref{fig:dots_200}} shows the alignment between $D_\mu \theta$ and $D_\mu \rho$ for all grid points in the large-scale simulation{, where we find} 98.5\% of grid cells show anti-alignment to within $\lesssim 1\%$. 
Figure~\ref{fig:dots_40} shows the same alignment in the simulation sampling {smaller-scale structures,} 
where we see a much larger spread of values across the domain. However, we still {find that} 66.9\% of grid cells show anti-alignment to within $<10\%$. 
We further note that we typically see anti-alignment in more under-dense regions, and alignment in areas surrounding over-dense regions.

In conclusion, our simulations are compatible with the constraints \eqref{brc1H}--\eqref{brc5H} for the properties that we have tested. In particular, the shear and electric Weyl curvature tensors have coinciding eigenbases. 
In addition, the stronger requirements $E_{\mu \nu} \propto \sigma_{\mu \nu}$ and $D_\mu \theta \propto D_\mu \rho$ apply to within {one} percent for most grid points in the present-epoch simulation domain for the large-scale simulation. 

}

\section{Cosmography for model-independent analysis of nearby sources}
\label{sec:modelindependent}
We now consider limits of the cosmographies \cite{Heinesen:2020bej,Heinesen:2021qnl} in the quiet universe approximation, based on our findings that it should provide a valid large-scale description of the universe. 

\subsection{Luminosity distance cosmography} 
We consider a generic congruence description of observers and emitters of light, and a space-time description within which luminosity distance, $d_L$, as a function of redshift $z$, is a well-defined function with a convergent Taylor series nearby the observer. 
Within such a description, the general cosmography is \cite{Heinesen:2020bej}  
\bea
\label{dLexpand}
    d_L  &=& d_L^{(1)} z   + d_L^{(2)} z^2 +  d_L^{(3)} z^3 + \mathcal{O}( z^4) \, , 
\eea 
where the coefficients can be expressed in the following way 
\bea
\label{dLexpand2}
    d_L^{(1)} &=& \frac{1}{\Eu_o} \, , \qquad d_L^{(2)} =   \frac{1 - \mathfrak{Q}_o }{2 \Eu_o}  \ , \qquad d_L^{(3)} =  \frac{- 1 +  3 \mathfrak{Q}_o^2 + \mathfrak{Q}_o    -  \mathfrak{J}_o   + \mathfrak{R}_o }{ 6  \Eu_o}     \, .
\eea 
The effective cosmological parameters $\Eu_o$, $\mathfrak{Q}_o$, $\mathfrak{R}_o$, and $\mathfrak{J}_o$ generalise the FLRW Hubble, $H_o$, deceleration, $q_o$, curvature, $\Omega_k$, and jerk, $j_o$, parameters, respectively, of the analogous luminosity distance cosmography in the FLRW limit \cite[see eq. (46) of][]{Visser:2003vq}. 
The generalised expressions (\ref{dLexpand2}) are necessary in order to consider local structure in space-time  
(and thus regional breaking of translational and rotational invariance) in a cosmographic treatment of observables.
The generalised cosmological parameters $\Eu_o$, $\mathfrak{Q}_o$, $\mathfrak{R}_o$, and $\mathfrak{J}_o$ can be formulated in terms of space-time kinematic variables and curvature DOFs as detailed in \cite{Heinesen:2020bej}\footnote{See also \cite{Umeh:2013UCT} for the first detailed derivation of the effective deceleration parameter.}. 

We shall now consider a congruence description of observers and emitters coinciding with the dust matter frame in the fluid model (\ref{def:dust}), (\ref{def:nablaH}). 
Setting {the acceleration} $a_\mu \equiv u^\nu \nabla_\nu u_\mu = 0$ {and the vorticity} $\omega_{\mu \nu} = 0$, as required by (\ref{def:dust}), 
{the} effective Hubble parameter of the cosmography (\ref{dLexpand}) reads 
\bea
\label{Eu}
    \Eu &=&  \frac{1}{3}\theta  +  e^\mu e^\nu \sigma_{\mu \nu}    \, ,
\eea  
where $\bm e$ is the spatial direction of the astrophysical source as seen on the observer's sky. The function $\Eu$ is a natural observed Hubble parameter, taking into account the inhomogeneity in expansion rate of space between observers, via spatial variations in $\theta$, and 
the anisotropy in the expansion rate over an individual observer's sky, through $\sigma_{\mu \nu}$. When evaluated at the observer, $\Eu_o$ {replaces}  
the Hubble constant in the observer's Hubble law. 
The effective deceleration parameter of the $d_L$ cosmography reads 
\bea
\label{q}
    \mathfrak{Q}(\bm e ) &=&  - 1 -  \frac{ \overset{0}{\mathfrak{q}}   +  \bm{e} \cdot  \bm{{\overset{1}{\mathfrak{q}}}}   +    \bm{e} \bm{e} \cdot  \bm{{\overset{2}{\mathfrak{q}}}}     +    \bm{e} \bm{e} \bm{e} \cdot  \bm{{\overset{3}{\mathfrak{q}}}}    +    \bm{e} \bm{e} \bm{e} \bm{e} \cdot  \bm{{\overset{4}{\mathfrak{q}}}}   }{\Eu^2(\bm e )}    \, , 
\eea 
where we have used the compact notation $\bm{e} \cdot  \bm{{\overset{1}{\mathfrak{q}}}} \equiv e^\mu  \overset{1}{\mathfrak{q}}_\mu$, $\bm{e} \bm{e}  \cdot  \bm{{\overset{2}{\mathfrak{q}}}} \equiv e^\mu e^\nu  \overset{2}{\mathfrak{q}}_{\mu \nu}$, and so on, and   
\bea
\label{qpoles}
    \hspace{-0.5cm} && \overset{0}{\mathfrak{q}} =   -\frac{1}{9}\theta^2  - \frac{1}{6}\kappa \rho   - \frac{11}{15} \sigma_{\mu \nu} \sigma^{\mu \nu}    \, , \qquad  \overset{1}{\mathfrak{q}}_\mu =     - \frac{3}{5} D_{\mu} \theta      \, , \nonumber \\
    \hspace{-0.5cm}  && \overset{2}{\mathfrak{q}}_{\mu \nu}  =   -  \frac{2}{3} \theta \sigma_{\mu \nu} - E_{\mu \nu} - \frac{13}{7} \sigma_{\alpha \la \mu} \sigma^\alpha_{\; \nu \ra }   \, , \qquad \overset{3}{\mathfrak{q}}_{\mu \nu \rho}  =  -  D_{ \la \mu} \sigma_{\nu   \rho \ra }      \, , \qquad       \overset{4}{\mathfrak{q}}_{\mu \nu \rho \kappa}  =   2   \sigma_{\la \mu \nu } \sigma_{\rho \kappa \ra} \, . 
\eea 
{In deriving the multipole coefficients \eqref{qpoles},} we have used \eqref{brc1H} and the geodesic deviation equation for $u^\rho \nabla_\rho \theta$ (the Raychaudhuri equation) and $u^\rho \nabla_\rho \sigma_{\mu \nu}$ \citep[see, e.g.,][]{Wald:106274}. 
All terms in the hierachy of multipoles (\ref{qpoles}) are given in terms of $\theta$ and $\sigma_{\mu \nu}$ ({i.e.,} the multipole components of $\Eu$), $\rho$, {$E_{\mu \nu}$, and} the spatial gradients of $\theta$ and $\sigma_{\mu \nu}$. 
Exploiting {the fact} that $E_{\mu \nu}$ and $\sigma_{\mu \nu}$ share eigenbases under the model ansatz (\ref{def:dust}), (\ref{def:nablaH}), $E_{\mu \nu}$ introduces only 2 additional DOFs (instead of 5 for a general traceless 2-component tensor of dimension 3), making the total number of independent DOFs determining $\mathfrak{Q}$ 13, instead of the general 16 {DOFs} \cite{Heinesen:2020bej}. 
{For the stronger condition $E_{\mu \nu} \propto \sigma_{\mu \nu}$, {which} {we} investigated in {Section~\ref{sec:propEsigma}}, the total number of independent DOFs introduced by $\mathfrak{Q}$ reduce further to 12. } 

The effective curvature parameter of the $d_L$ cosmography reads 
\bea
\label{R}
\mathfrak{R} (\bm e ) &=&  -  \frac{  \overset{0}{\mathfrak{r}}   +  \bm{e} \cdot  \bm{{\overset{1}{\mathfrak{r}}}}   +    \bm{e} \bm{e} \cdot  \bm{{\overset{2}{\mathfrak{r}}}}     +    \bm{e} \bm{e} \bm{e} \cdot  \bm{{\overset{3}{\mathfrak{r}}}}    +    \bm{e} \bm{e} \bm{e} \bm{e} \cdot  \bm{{\overset{4}{\mathfrak{r}}}}        }{\Eu^2(\bm e )}  \, , 
\eea 
with coefficients
\bea
\label{Rpoles} 
&& \overset{0}{\mathfrak{r}} =  \overset{0}{\mathfrak{q}} + \frac{1}{2} \kappa \rho   \, , \qquad  \overset{1}{\mathfrak{r}}_\mu =   \overset{1}{\mathfrak{q}}_{\mu}     \, , \qquad \overset{2}{\mathfrak{r}}_{\mu \nu}  =     \overset{2}{\mathfrak{q}}_{\mu \nu}   \,    , \qquad     \overset{3}{\mathfrak{r}}_{\mu \nu \rho}  =    \overset{3}{\mathfrak{q}}_{\mu \nu \rho}  \, ,   \qquad  \overset{4}{\mathfrak{r}}_{\mu \nu \rho \kappa}  =    \overset{4}{\mathfrak{q}}_{\mu \nu \rho \kappa}  \, , 
\eea 
where we have used (\ref{def:dust}) along with Einstein's field equations to relate the Ricci curvature of the space-time to the energy momentum content. 
We see that the anisotropies of $\mathfrak{R}$ are fully determined by the multipole coefficients of $\Eu$ and $\mathfrak{Q}$, due to the absence of anisotropic stresses and flux of energy in the quiet universe model. 
The effective curvature parameter thus does not introduce {any} additional DOFs under our model assumptions. 

{We calculate} the {simplified} effective jerk parameter from its exact multipole decomposition given in Appendix~B of \cite{Heinesen:2020bej}. 
{Due to its lengthy expression, 
{we show} {the simplified} decomposition of $\mathfrak{J}$ 
in {Appendix~\ref{sec:jerk}}. 
Combining (\ref{jpoles}) and (\ref{qpoles}), we {find} that the $2^5$-pole of the jerk parameter seris expansion,} $\overset{5}{\mathfrak{j}}_{\mu \nu \rho \kappa \gamma}$, is completely specified by $\sigma_{\mu \nu}$ and $\overset{3}{\mathfrak{q}}_{\mu \nu \rho}$, thus reducing the 
DOFs introduced by $\mathfrak{J}$ from 36 to 25.

We can further reduce the DOFs by considering the dominant {anisotropic contributions} in the hierarchy of multipoles, which for most observers in realistic universe models are expected to be {those} containing a maximum number of spatial gradients of kinematic variables \citep[see][for a discussion on dominant multipoles for typical observers]{Macpherson:2021a}. 
For $\mathfrak{Q}$, this is the dipole $\overset{1}{\mathfrak{q}}_\mu$ (containing a spatial derivative of $\theta$) and the octupole $\overset{3}{\mathfrak{q}}_{\mu \nu \rho}$ (containing a spatial derivative of $\sigma_{\mu \nu}$),
{which are also} the multipoles {dominating $\mathfrak{R}$.} 
The effective jerk parameter, $\mathfrak{J}$, is dominated by $\overset{2}{\mathfrak{j}}_{\mu \nu}$ (containing second order spatial derivatives of $\theta$ and $\sigma_{\mu \nu}$) and $\overset{4}{\mathfrak{j}}_{\mu \nu \rho \kappa}$ (containing second order spatial derivatives $\sigma_{\mu \nu}$). 
{Accounting only for the dominant multipoles, $\mathfrak{Q}$ is specified by 11 DOFs, whereas $\mathfrak{J}$ is specified by 15 independent DOFs. The effective curvature parameter, $\mathfrak{R}$, is fully determined from the multipoles of $\Eu$, $\mathfrak{Q}$, and $\mathfrak{J}$.   } 

{The total number of DOFs specifying the third order luminosity distance cosmography {under our approximations is} 32 (as reduced from 61 DOFs in the most general case).}

{Finally, we} pay particular attention to the dipolar signature of the effective cosmological parameters. The effective Hubble parameter, $\Eu$, has no dipolar signature, since its only anisotropic feature is a quadrupolar term {for geodesic observers}.
Under the quiet universe approximation, the dipole of the deceleration parameter is $\overset{1}{\mathfrak{q}}_{\mu}\propto D_\mu \theta$. 
For our large-scale simulations, 
{we find} $D_\mu \theta \propto D_\mu \rho$ {to a good approximation} {(see Figure~\ref{fig:dots_200})}, and the dipole of the deceleration parameter {will thus} 
be directed along the axis defined by the {spatial} gradient of the {local} density field. } 
The dipole term of {the jerk parameter,} 
$\overset{1}{\mathfrak{j}}_{\mu}$, is dominated by terms proportional to $D_\mu \overset{0}{\mathfrak{q}}$ and $D^{\nu} \overset{2}{\mathfrak{q}}_{\nu \mu}$ (see Appendix~\ref{sec:jerk}). 
Neglecting terms which are second order in shear in \eqref{qpoles}, and evaluating the derivatives under our model assumptions, we {arrive at} 
$\overset{1}{\mathfrak{j}}_{\mu}\propto D_\mu \theta \propto D_\mu \rho$. 
{The} dipole of $\mathfrak{J}$ thus aligns with the dipole of $\mathfrak{Q}$. 

The dipolar feature {as predicted by the cosmography} is interesting in light of dipoles detected in distance--redshift data \cite{Colin:2019opb,Migkas:2021zdo} and in other cosmological probes \citep[see Figure~22 and Table~IV of][]{Perivolaropoulos:2021jda}, which {are found to be} approximately aligned with the CMB dipole.

\subsection{Redshift drift cosmography} 
{
{Here we simplify the general cosmography for analysing redshift drift signals,  formulated in}
\cite{Heinesen:2021qnl}, {under the quiet universe assumption}. 
As discussed in \cite{Heinesen:2021qnl}, the cosmography for redshift drift involves information on the \emph{position} drift {$\bm \kappa$} of the source (together with the position of the source itself), which complicates the model-independent expressions for the redshift drift signal. 
Therefore, we will analyse only the first order term in the series expansion, namely 
\bea
\label{zdfirst2}
\frac{d z}{d \tau} \Bigr\rvert_{\obs} =  - \mathbb{Q}_\obs \Eu_\obs z + \mathcal{O}(z^2)  \, ,  
\eea  
where the effective deceleration parameter 
is given by 
\bea
\label{Pimultikappa}
    \hspace*{-0.65cm} \mathbb{Q} = - \frac{ -  \kappa^\mu \kappa_\mu  + \Sigma^{\it{o}}    +  e^\mu \Sigma^{\bm{e}}_\mu    +       e^\mu   e^\nu \Sigma^{\bm{ee}}_{\mu \nu} + e^\mu   \kappa^\nu \Sigma^{\bm{e\kappa}}_{\mu \nu} }{\Eu^2} \, .
\eea  
{Under} 
the model assumptions \eqref{def:dust} and \eqref{def:nablaH}, 
{the coefficients of $\mathbb{Q}$} reduce to 
\bea
\label{Picoefkappa}
    &&\hspace{-0.5cm} \Sigma^{\it{o}} =    - \frac{1}{6} \kappa \rho        \, , \; \;  \Sigma^{\bm{e}}_\mu  =    0 \, ,  \; \;  \Sigma^{\bm{ee}}_{\mu \nu} =    -  E_{\mu \nu}    \,   , \; \;  \Sigma^{\bm{e\kappa}}_{\mu \nu} =   2 \sigma_{\mu \nu}    \, .
\eea  
The first order redshift drift cosmography is very simple in its form: it contains the DOFs $\theta$ and $\sigma_{\mu \nu}$ {inherited from}  
$\Eu$ \eqref{Eu}, and 
the independent DOFs 
{from} $\rho$ and $E_{\mu \nu}$ entering the coefficients of the effective deceleration parameter \eqref{Picoefkappa}. 
Under {the quiet universe} {approximation}, the eigenbasis of $E_{\mu \nu}$ is the same as that of $\sigma_{\mu \nu}$, and $E_{\mu \nu}$ {thus} introduces two 
independent scalar DOFs -- or one independent DOF when the stronger condition $E_{\mu \nu} \propto \sigma_{\mu \nu}$ applies (see Section~\ref{sec:propEsigma}).  

{As argued in \cite{Heinesen:2021qnl}, the last term in the numerator of \eqref{Pimultikappa} might be considered as a second order term for realistic modelling, due to the expected position drift signals $\bm \kappa$ being of much smaller amplitude than the local expansion rate for observations made at cosmological scales. 
Thus, we expect the quadrupole, $\Sigma^{\bm{ee}}_{\mu \nu} = -  E_{\mu \nu}$, to dominate the anisotropic signature of low-redshift measurements of redshift drift{, together with the quadrupole, $\sigma_{\mu \nu}$, entering in the denominator of \eqref{Pimultikappa}. When the proportionality law $E_{\mu \nu} \propto \sigma_{\mu \nu}$ holds, these two quadrupolar contributions to the redshift drift signal are proportional.}  }  
{In this most simplified case, the first order redshift drift signal is given by 8 DOFs in total ({2 DOFs} introduced by $\rho$ and $E_{\mu \nu}$ in $\mathbb{Q}$ in addition to the 6 DOFs {in} $\Eu$), in comparison to 21 DOFs in the most general case. }

As noted in \cite{Heinesen:2021qnl}, the effective deceleration parameter of the redshift drift cosmography, $\mathbb{Q}$, is distinct from  
the deceleration parameter of the luminosity distance cosmography, $\mathfrak{Q}$. 
In particular, $\mathfrak{Q}$ contains spatial gradients of the kinematic variables of the observer congruence whereas $\mathbb{Q}$ is \sayy{blind} to such spatial gradients. As a consequence, we expect the amplitude of the anisotropic signal in $\mathbb{Q}$ to be lower than for $\mathfrak{Q}$. 
}

\section{Discussion and conclusions} 
\label{sec:conclusion} 
{We} have examined the applicability of the quiet universe approximation {\cite{Maartens:1996uv,Sopuerta:1998rt}} in realistic large-scale cosmological simulations {evolved using numerical relativity. The quiet universe class of models accurately} captures the physics of the simulations, {confirming that these models are useful to describe the large-scale universe within general relativity.}

{We used the quiet universe to simplify two fully general cosmographic expansions, thus providing predictions for} the anisotropic features in luminosity distance and redshift drift signals in this model limit. 
The number of DOFs describing the cosmographies {are reduced} significantly, especially for the redshift drift signal, with an anisotropic signature which 
is dominated by a quadrupolar term given by the electric Weyl curvature tensor.
Considering only the leading order multipoles of the luminosity distance cosmography  further reduces the number of DOFs involved for this observable. 
{In the most simplified versions of the cosmographies that we consider, the number of DOFs specifying the third order luminosity distance cosmography reduce to 32 (from 61 DOFs in the general case), whereas the number of DOFs specifying the first order redshift drift signal reduce to 8 DOFs (from 21 DOFs).} 

Based on the quiet universe approximation and the approximate alignment $D_\mu \theta \propto D_\mu \rho$ found in our simulated large scale universe, we {further} predict that the dipolar feature {in the luminosity distance} at low redshifts is aligned with the spatial gradient of density, $D_\mu \rho$, as evaluated at the observer. 
\emph{Consequently, we predict a dipolar signature in the distance-redshift relation for low redshift data of standardisable objects that is aligned with the gradient of the large scale density field.} 
We stress that the signature of this prediction can in general not be accounted for by a pointwise special-relativistic boost of the observer. However, coherent bulk flow motions can create multipole signatures in distance-redshift cosmography 
\cite{2001AstL...27..765P}, which {for certain peculiar flow models} might resemble those {we predict here.} 

{As remarked in \cite{Clarkson:2010uz}, for the luminosity distance--redshift relation, the anisotropy of observables are tightly linked to anisotropies in space-time geometry. In \emph{any} universe with structure, cosmological observables will necessarily be anisotropic over the observers' skies. For instance, an everywhere isotropic effective Hubble parameter \eqref{Eu} requires the shear tensor to vanish everywhere. For the irrotational dust space-times considered here, this immediately implies that the geometry is exactly FLRW \cite{Ellis:2011pi}. 
{It is therefore} clear that observables like luminosity distance and redshift drift signals will be anisotropic over our sky, {however}, the signatures and \emph{amplitude} of these anisotropies must be tested with data. } 

As a byproduct of our analysis, we {have also shown} that the silent universe approximation{---}a restricted class of quiet universe models without the propagation of gravitational waves{---}fails to capture the physics of our large-scale cosmological {simulations}. Thus, gravitational radiation as covariantly quantifed through the magnetic Weyl curvature tensor, 
even though small in amplitude, has important implications for large-scale cosmological modelling. We find this non-trivial insight valuable for future accurate modelling of cosmological dynamics and large-scale structure. 

{In conclusion, our main findings can be summarised as follows:
\begin{itemize}
    \item The \sayy{quiet universe} approximation accurately describes the physics of large-scale cosmological simulations performed with numerical relativity
    \item We predict a dipolar signature in low-redshift {luminosity distances} 
    which is aligned with the gradient of local density contrasts
    \item {We predict the lowest order redshift drift signal to be dominated by a quadruple feature, aligned with the shear tensor and the electric Weyl tensor as evaluated at the observer} 
    \item The silent universe approximation \emph{does not} provide a good description of our large-scale simulations, emphasising the potential importance of including gravitational radiation in cosmological modelling 
\end{itemize}
} 

{We remark that the properties of the quiet universe models as described in Section~\ref{sec:theory} carry over to space-times that include a cosmological constant. Thus, our main conclusions are expected to hold in the presence of a cosmological constant or another dark energy-type component {with homogeneous pressure}. }
{The reduced cosmographic framework presented in this paper 
{is useful} for the analysis of {upcoming large} distance--redshift catalogues as well as 
{future} measurements of redshift drift signals. 
Our results give direct predictions for the expected anisotropic signatures in these observables.  }

\vspace{6pt} 
\begin{acknowledgments}
We would like to thank Thomas Buchert and Roy Maartens for helpful comments on the manuscript. 
This work is part of a project that has received funding from the European Research Council (ERC) under the European Union's Horizon 2020 research and innovation programme (grant agreement ERC advanced grant 740021--ARTHUS, PI: Thomas Buchert). HJM appreciates support from the Herchel Smith Postdoctoral Fellowship fund. The simulations in this work used the DiRAC@Durham facility managed by the Institute for Computational Cosmology on behalf of the STFC DiRAC HPC Facility (www.dirac.ac.uk). The equipment was funded by BEIS capital funding via STFC capital grants ST/P002293/1, ST/R002371/1 and ST/S002502/1, Durham University and STFC operations grant ST/R000832/1. DiRAC is part of the National e-Infrastructure.

\end{acknowledgments}

\appendix

\section{Richardson extrapolation of errors}

As mentioned in the main text, we perform three simulations with identical initial data so that we can perform a Richardson extrapolation and quantify our error bars on the present-epoch slice. 
We do this only for the simulation with structure beneath 200$h^{-1}$Mpc removed from the initial data, because simulations containing any smaller-scale structure will develop different physical gradients at $z=0$ between resolutions and thus cannot be compared using these methods.

The Richardson extrapolation is based on the assumption that our numerical estimates from the simulations will approach the ``true'' values of the physical quantities as we increase our numerical resolution $N\rightarrow\infty$. The rate at which we approach the true value depends on the accuracy of the numerical scheme used. Here all of our calculations are fourth-order accurate, implying our numerical estimates should approach the ``true'' solution at a rate $\propto 1/N^4$. 

We estimate the {error of the Weyl and shear scalars \eqref{eq:Eandshear}} by calculating them at all points in the three simulations with $N=64,128$, and 256. For each shared coordinate point, we then fit a curve of the form $f(N) = a + b/N^4$ to the three values of $N$, where $a$ and $b$ are parameters we determine using the \texttt{curve\_fit} function in the \texttt{SciPy}\footnote{\url{https://scipy.org}} package. Extrapolating the determined function to very large $N$, here we take $N=10^5$, gives an estimate of the ``true'' value of that quantity. The error at the highest resolution, $N=256$, is thus determined as the relative difference between the numerical and the extrapolated ``true'' value, e.g. for the Weyl scalar
\begin{equation}
    {\rm error}(E) \equiv \frac{E_{N=256}}{E_{{\rm extrap},N=10^5}} - 1,
\end{equation}
and similarly for the shear scalar $\sigma$.

\begin{figure*}[!h]
    \centering
    \includegraphics[width=0.8\textwidth]{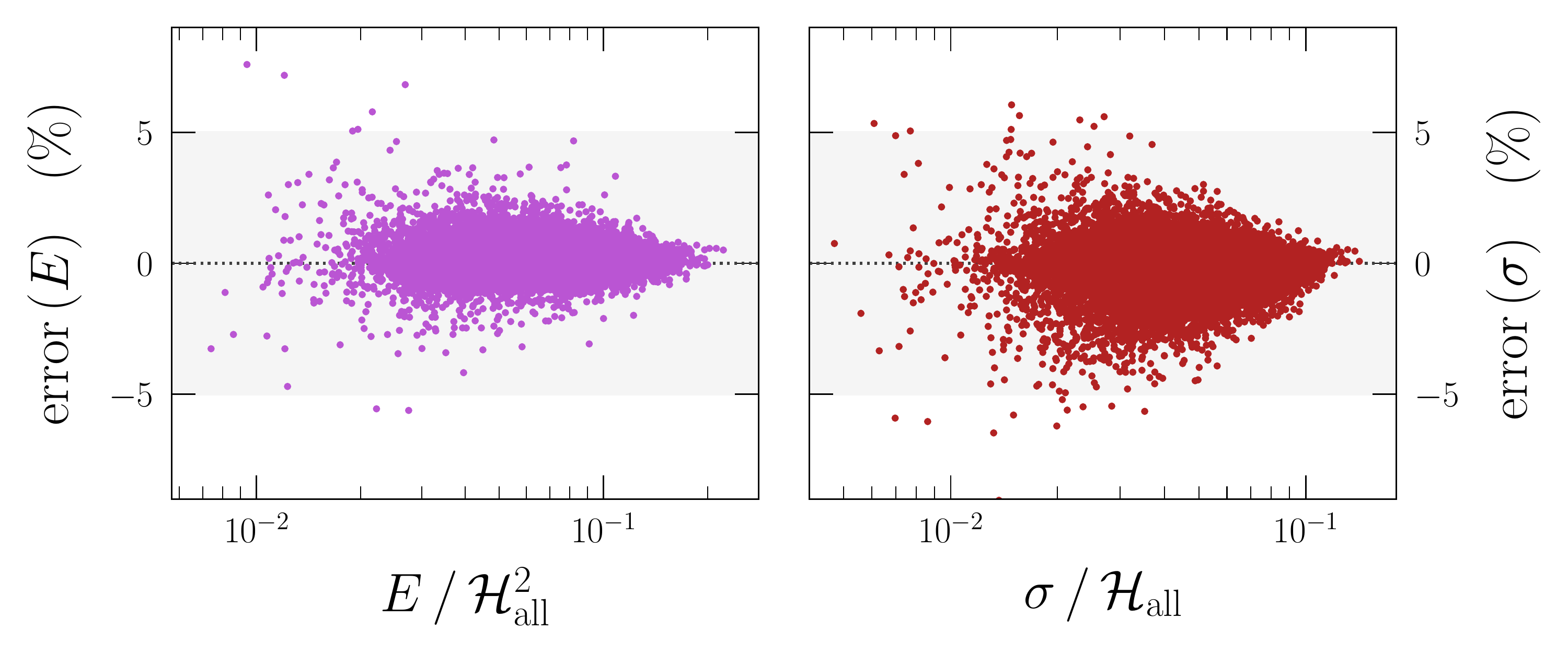}
    \caption{Richardson extrapolated error in the electric Weyl scalar $E$ (left panel) and in the shear scalar $\sigma$ (right panel). Points show the error at half of the grid cells in the $N=256$ simulation shared with lower-resolution runs, as a function of the value of ${E}/\mathcal{H}^2_{\rm all}$ or $\sigma/\mathcal{H}_{\rm all}$ at that point. The shaded band shows $\pm 5\%$ errors for reference.}
    \label{fig:Esigrich}
\end{figure*}
Figure~\ref{fig:Esigrich} shows the percentage error for the Weyl (left panel) and shear (right panel) scalars, as a function of the value of each respective quantity at that coordinate point (normalised by the globally-averaged Hubble parameter). We show the error at half of the grid cells in the $N=256$ simulation shared with the lower-resolution simulations, i.e. $(64/2)^3$ grid cells in total, with a $\pm 5\%$ grey shaded region for reference. The errors in the Weyl (shear) scalar are less than $2\%$ for 99.5 (97.6) \% of shared grid points.


\section{The effective jerk parameter} 
\label{sec:jerk}
We consider the effective jerk parameter, which reads 
\bea
\label{jerk}
\mathfrak{J}(\bm e ) &=& 1 +   \frac{  \overset{0}{\mathfrak{j}}   +  \bm{e} \cdot  \bm{{\overset{1}{\mathfrak{j}}}}   +    \bm{e} \bm{e} \cdot  \bm{{\overset{2}{\mathfrak{j}}}}     +    \bm{e} \bm{e} \bm{e} \cdot  \bm{{\overset{3}{\mathfrak{j}}}}    +    \bm{e} \bm{e} \bm{e} \bm{e} \cdot  \bm{{\overset{4}{\mathfrak{j}}}}      +   \bm{e} \bm{e} \bm{e} \bm{e} \bm{e} \cdot  \bm{{\overset{5}{\mathfrak{j}}}}     +   \bm{e} \bm{e} \bm{e} \bm{e} \bm{e} \bm{e} \cdot  \bm{{\overset{6}{\mathfrak{j}}}}  }{\Eu^3(\bm e )}    
\eea 
with coefficients 
\bea
\label{jpoles}
&& \overset{0}{\mathfrak{j}}  =   \frac{ {\rm d} \overset{0}{\mathfrak{q}}  }{{\rm d} \tau} + \theta \overset{0}{\mathfrak{q}}   - \frac{1}{3} D^{ \mu} \overset{1}{\mathfrak{q}}_{\mu }    - \frac{2}{3} \sigma^{ \mu \nu}   \overset{2}{\mathfrak{q}}_{\mu \nu}     + h^{\mu \nu}   h^{\rho \kappa}   \sigma_{ ( \mu \nu } \overset{2}{\mathfrak{q}}_{\rho \kappa )}      \, , \nonumber \\
&&  \overset{1}{\mathfrak{j}}_\mu =  - D_\mu \overset{0}{\mathfrak{q}}  + h^{\nu}_{\, \mu} \frac{ {\rm d}  \overset{1}{\mathfrak{q}}_{\nu}  }{{\rm d} \tau}  +  \theta \overset{1}{\mathfrak{q}}_{\mu}  - \sigma^\nu_{\; \mu} \overset{1}{\mathfrak{q}}_{\nu}  
      +  \frac{18}{7}  h^{\nu \rho}   h^{\kappa \gamma}    \sigma_{( \mu \nu }  \overset{3}{\mathfrak{q}}_{ \rho \kappa \gamma)}   \nonumber \\ 
  && \qquad \quad + \frac{1}{5} h^{\nu \rho} \left( 4  \sigma_{ (\mu \nu } \overset{1}{\mathfrak{q}}_{\rho )}   - D_{(\mu} \overset{2}{\mathfrak{q}}_{\nu \rho)}        -3  \sigma^\kappa_{\; ( \mu}  \overset{3}{\mathfrak{q}}_{\nu \rho ) \kappa } \right)       \, , \nonumber \\
 && \overset{2}{\mathfrak{j}}_{\mu \nu}  =    3  \overset{0}{\mathfrak{q}} \sigma_{\mu \nu}   - D_{ \la \mu} \overset{1}{\mathfrak{q}}_{\nu \ra }    - 2 \sigma^\rho_{\; \la \mu}   \overset{2}{\mathfrak{q}}_{\nu \ra \rho}    + h^{\alpha}_{\, \mu} h^{\beta}_{\, \nu } \frac{ {\rm d}  \overset{2}{\mathfrak{q}}_{\alpha \beta} }{{\rm d} \tau}  +   \theta \overset{2}{\mathfrak{q}}_{\mu \nu}            \nonumber \\ 
 && \qquad \quad +   \frac{6}{7} h_{\la \mu}^{\, \alpha} h_{\nu \ra}^{\, \beta}   h^{\rho \kappa}  \left(  5 \sigma_{ ( \alpha \beta } \overset{2}{\mathfrak{q}}_{\rho \kappa )}     - D_{(\alpha}  \overset{3}{\mathfrak{q}}_{\beta \rho \kappa)}    - 4 \sigma^\gamma_{\; ( \alpha}  \overset{4}{\mathfrak{q}}_{\beta \rho \kappa ) \gamma  }        \right)    +   5  h_{\la \mu}^{\, \alpha} h_{\nu \ra}^{\, \beta}   h^{\rho \kappa}  h^{\gamma \sigma} \sigma_{( \alpha \beta }  \overset{4}{\mathfrak{q}}_{ \rho \kappa \gamma \sigma ) } \, ,   \nonumber \\ 
&&  \overset{3}{\mathfrak{j}}_{\mu \nu \rho}  =     4  \sigma_{ \la \mu \nu } \overset{1}{\mathfrak{q}}_{\rho \ra }   - D_{\la \mu} \overset{2}{\mathfrak{q}}_{\nu \rho \ra }      -3  \sigma^\kappa_{\; \la \mu}  \overset{3}{\mathfrak{q}}_{\nu \rho \ra  \kappa }           +  h^{\alpha}_{\, \mu} h^{\beta}_{\, \nu}   h^{\gamma}_{\, \rho}    \frac{ {\rm d}  \overset{3}{\mathfrak{q}}_{\alpha \beta \gamma}   }{{\rm d} \tau}   + \theta  \overset{3}{\mathfrak{q}}_{\mu \nu \rho}   \nonumber \\  
&& \qquad \quad    + \frac{10}{9} h_{\la \mu}^{\, \alpha} h_\nu^{\, \beta} h_{\rho \ra}^{\, \epsilon}   h^{\kappa \gamma}  \left(  6   \sigma_{( \alpha \beta }  \overset{3}{\mathfrak{q}}_{ \epsilon \kappa \gamma)} - D_{( \gamma} \overset{4}{\mathfrak{q}}_{\alpha \beta \epsilon \kappa ) }   \right)   \, , \nonumber \\ 
&&   \overset{4}{\mathfrak{j}}_{\mu \nu \rho \kappa}  =   5 \sigma_{ \la \mu \nu } \overset{2}{\mathfrak{q}}_{\rho \kappa \ra }  - D_{\la \mu}  \overset{3}{\mathfrak{q}}_{\nu \rho \kappa \ra}   - 4 \sigma^\gamma_{\; \la \mu}  \overset{4}{\mathfrak{q}}_{\nu \rho \kappa \ra \gamma  }   \nonumber  \\ 
&&\qquad \quad +  h^{\alpha}_{\,  \mu} h^{\beta}_{\, \nu  } h^{\sigma}_{\, \rho} h^{\eta}_{\, \kappa}   \frac{ {\rm d}    \overset{4}{\mathfrak{q}}_{\alpha \beta \sigma \gamma}   }{{\rm d} \tau}   + \theta \overset{4}{\mathfrak{q}}_{\mu \nu \rho \kappa}  + \frac{105}{11} h_{\la \mu}^{\, \alpha} h_\nu^{\, \beta} h_\rho^{\, \epsilon} h_{\kappa \ra}^{\, \psi}   h^{\gamma \sigma} \sigma_{( \alpha \beta }  \overset{4}{\mathfrak{q}}_{ \epsilon \psi \gamma \sigma ) }      \, , \nonumber \\ 
  &&   \overset{5}{\mathfrak{j}}_{\mu \nu \rho \kappa \gamma}  =    6   \sigma_{\la \mu \nu }  \overset{3}{\mathfrak{q}}_{ \rho \kappa \gamma \ra} - D_{\la \gamma} \overset{4}{\mathfrak{q}}_{\mu \nu \rho \kappa \ra }    \, , \nonumber \\ 
&&   \overset{6}{\mathfrak{j}}_{\mu \nu \rho \kappa \gamma \sigma}  =      7 \sigma_{\la \mu \nu }  \overset{4}{\mathfrak{q}}_{ \rho \kappa \gamma \sigma \ra } \, .
\eea 
The multipole coefficients of $\mathfrak{J}$ are determined fully from the multipole coefficients of $\mathfrak{Q}$ and their first derivatives, together with $\theta$ and $\sigma_{\mu \nu}$. 


\bibliographystyle{plain}
\bibliography{CosmographyQuiet}{}


\end{document}